\documentclass[preprint,superscriptaddress,nofootinbib]{revtex4-1}
\usepackage{hyperref}
\usepackage{bbm}
\usepackage{amsfonts}
\usepackage{mathrsfs}

\usepackage{amsmath,amssymb}

\usepackage{epsfig}
\usepackage{graphicx}               % Standard graphics package
\usepackage{url}
\usepackage{hyperref}
\usepackage{float}
\usepackage{pstricks}
\usepackage{color}

\newcommand{\nc}{\newcommand}

\nc{\beq}{\begin{equation}}
\nc{\eeq}{\end{equation}}
\nc{\beqa}{\begin{eqnarray}}
\nc{\eeqa}{\end{eqnarray}}
\nc{\bea}{\begin{eqnarray}}
\nc{\eea}{\end{eqnarray}}
\nc{\barray}{\begin{eqnarray}}
\nc{\earray}{\end{eqnarray}}
\nc{\barrayn}{\begin{eqnarray*}}
\nc{\earrayn}{\end{eqnarray*}}
\nc{\ra}{\rightarrow}
\newcommand{\lsim}{\!\mathrel{\hbox{\rlap{\lower.55ex \hbox{$\sim$}} \kern-.34em \raise.4ex \hbox{$<$}}}}
\newcommand{\gsim}{\!\mathrel{\hbox{\rlap{\lower.55ex \hbox{$\sim$}} \kern-.34em \raise.4ex \hbox{$>$}}}}
\nc{\Tr}{{\rm Tr}}
\nc{\slsh}{\slash\hspace*{-0.22cm}}
\def\eg{{\it e.g.}}
\def\ie{{\it i.e.}}
\def\be{\begin{equation}}
\def\ee{\end{equation}}
\def\bea{\begin{eqnarray}}
\def\eea{\end{eqnarray}}

\newcommand{\missET}{\slash{\hspace{-2.5mm}E}_T}
\nc{\infinity}{\infty}
\nc{\mc}{\mathcal}
\nc{\M}{\mathcal{M}}

\newcommand{\met}{E\!\!\!/_T}

\newcommand{\sth}{{\tilde{t}_2}}
\newcommand{\stl}{{\tilde{t}_1}}
\newcommand{\sbot}{{\tilde{b}_1}}
\newcommand{\msth}{{m_{\tilde{t}_2}}}
\newcommand{\mstl}{{m_{\tilde{t}_1}}}
\newcommand{\msbot}{{m_{\tilde{b}_1}}}

\def\ie{{\it i.e.}}
\def\eg{{\it e.g.}}

\def\to{\rightarrow}

\begin{document}

\title{Implications of  a Stop Sector Signal at the LHC}

\author{Aaron Pierce}
\affiliation{Michigan Center for Theoretical Physics, \\University of Michigan, Ann Arbor, MI 48109, USA}
\author{Bibhushan Shakya}
\affiliation{Department of Physics, University of Cincinnati, Cincinnati, OH 45221, USA}
\affiliation{Leinweber Center for Theoretical Physics, Department of Physics\\University of Michigan, Ann Arbor, MI 48109, USA}

\begin{abstract}
Naturalness arguments suggest that the stop sector is within reach of the Large Hadron Collider (LHC).  We investigate how the observation of a third generation squark signal could  predict  masses and discovery modes of other supersymmetric particles, or potentially test the Higgs boson mass relation and the validity of the Minimal Supersymmetric Standard Model (MSSM) at the high luminosity LHC. We illustrate these ideas in three distinct scenarios:  discovery of a light stop,  a sbottom signal in multileptons, and a signal of the second (heavier) stop in boosted dibosons.
\end{abstract}

\preprint{MCTP-16-25}

\maketitle

%\tableofcontents

%%%%%%%%%%%%%%%%%%%%%%%%%%%%%%%%
\section{Introduction and Motivation}
\label{sec:introduction}
%%%%%%%%%%%%%%%%%%%%%%%%%%%%%%%%

If the electroweak scale is natural,  third generation squarks should be among the first supersymmetric particles to be discovered at the Large Hadron Collider (LHC).  The latest results from Run II of the LHC place strong limits on their masses under various assumptions about the mass spectrum and decay channels \cite{Aaboud:2016lwz,Aad:2015pfx,Khachatryan:2016pxa,CMS-PAS-SUS-16-007,Khachatryan:2016kod,ATLAS-CONF-2016-077,ATLAS-CONF-2016-076,CMS-PAS-SUS-16-029,Aaboud:2016tnv,Aaboud:2017nfd,Aaboud:2017ayj,ATLAS-CONF-2017-037,Sirunyan:2017wif,Sirunyan:2017xse,Sirunyan:2017leh,Sirunyan:2017kiw}. Nevertheless, windows for light  (\,$\lsim$ TeV) third generation squarks still exist, and there have even been recent hints of signals of such light states (e.g. \cite{Collins:2015boa,Huang:2015fba}).  The discovery of a third generation squark at the LHC in the next few years remains an exciting possibility.

Such a discovery carries important implications, both theoretical and observational. One of the primary appeals of a stop sector discovery is that it is intricately tied to the mass of the Higgs boson \cite{Ellis:1990nz,Haber:1990aw,Okada:1990vk}.  Given the measurements of the mass and properties of the Higgs in recent years, this connection provides strong constraints on the possible values of stop masses and mixing, which, in turn, determine their decay branching ratios (see \eg\,\cite{Guo:2013iij}). Furthermore, the left-handed stop is part of a doublet that also contains the left-handed sbottom, hence their masses are related: in particular, after mixing in the stop sector, the left-handed sbottom mass lies between the two stop mass eigenstates provided sbottom mixing is not too large. Such correlations imply that an initial signal can enable predictions of subsequent signals at the LHC. Establishing discrepancies between the observed Higgs mass and that predicted from third generation sparticle measurements could rule out the Minimal Supersymmetric Standard Model (MSSM) as the underlying theory behind these signals, establishing the need for a non-minimal version of supersymmetry, such as the Next-to-Minimal Supersymmetric Standard Model (NMSSM). Such predictions and consistency checks can remain largely insensitive to the remainder of the supersymmetric mass spectrum. 

In this paper, we study such theoretical and observational implications of a stop sector signal at the LHC within a few specific scenarios. These are not intended to provide comprehensive coverage of all possibilities, but rather offer qualitative illustrations of the various ways in which progress can be made once a signal is observed.\,\footnote{Earlier ideas using measurements to constrain parameters in the stop sector include \cite{Carena:1998wq,Rolbiecki:2009hk}.} In the event of a relevant discovery of the type described here, it would be of interest to carry out the corresponding theory calculations with higher precision (i.e. higher loop level) and examine the collider aspects (i.e. event and background simulations) with additional care.

Finally, we elaborate on the philosophy behind the structure of this paper. MSSM parameter space studies generally scan over all parameters in the theory over some range, calculate the Higgs mass at two or three loops, and include all relevant constraints from flavor, dark matter, and other relevant aspects. While we also scan over stop parameters in this paper, our setup is manifestly different. Our studies are driven by hypothetical \textit{observations}:  in particular, we are interested in scenarios where stop parameters are known with some uncertainty due to observed signals, but other parameters in the underlying theory, such as the gluino mass, are not known at all. Then it becomes impossible to calculate the Higgs mass at higher order, and we instead allow the Higgs mass at one loop within a reasonable window that includes all potentially consistent regions of parameter space (see Sec.\,\ref{sec:framework} for details). Likewise, given the lack of information on other parameters, we also do not include any constraints from flavor, dark matter, or other similar considerations that rely on additional assumptions or parameters not relevant to our study of the stop sector. A proper inclusion of such constraints or the calculation of the Higgs mass with greater precision would eliminate a subset of the parameter space points we consider in various sections in this paper; however, this would not falsify any of the statements or conclusions in these sections, but only make them sharper and stronger. 

The paper is structured as follows.  The basic theoretical framework and relevant observational constraints are reviewed in Sec.\,\ref{sec:framework}. In the next three sections, we study distinct scenarios where the Higgs mass relation can be used to perform consistency checks of the MSSM framework, or alternatively predict masses and decay modes of other superpartners. In Sec.\,\ref{sec:stop1}, we investigate how predictions can be made for sbottom or heavier stop masses and decay modes following a light stop discovery. Sec.\,\ref{sec:sbottom} investigates how information about the heavier stop can be deduced from measurements of a sbottom signal in multileptons; in this case, this signal reveals information about multiple parameters in the stop sector, enabling very precise predictions about the heavier stop, aiding in its discovery -- we illustrate this with a benchmark case study. In Sec.\,\ref{sec:stop2}, we study how measurements of the heavier stop decay in boosted diboson channels, which carry information about multiple stop sector parameters, can be used to test the consistency of the MSSM Higgs mass relation and either corroborate or rule out the MSSM; this section is also supplemented with a benchmark case study.

%%%%%%%%%%%%%%%%%%%%%%%%%%%%%%%%
\section{Theoretical Framework}
\label{sec:framework}
%%%%%%%%%%%%%%%%%%%%%%%%%%%%%%%%
We denote the lighter and heavier stop mass eigenstates as $\stl$ and $\sth$ respectively. We denote the stop mixing angle as $\theta_t$, with $\stl=\cos\theta_t\, \tilde{t}_L + \sin\theta_t \,\tilde{t}_R$, where $\tilde{t}_L ,\,\tilde{t}_R$ are the stop gauge eigenstates, so that $\theta=0$ corresponds to the scenario where the lighter stop is left-handed. In terms of these parameters, the Higgs boson mass at one-loop in the MSSM is \cite{Martin:1997ns}
\beqa
m_h^2&=&m_Z^2\cos^2(2\beta)+\frac{3 \sin^2\beta\, y_t^2}{4\pi^2}\, [\, m_t^2\ln\left(\frac{m_{\tilde{t}_1}m_{\tilde{t}_2}}{m_t^2}\right)+c_t^2s_t^2(m_{\tilde{t}_2}^2-m_{\tilde{t}_1}^2)\ln\left(\frac{m_{\tilde{t}_2}^2}{m_{\tilde{t}_1}^2}\right)\nonumber\\
&+&c_t^4s_t^4\left\{(m_{\tilde{t}_2}^2-m_{\tilde{t}_1}^2)^2-\frac{1}{2}(m_{\tilde{t}_2}^4-m_{\tilde{t}_1}^4)\ln\left(\frac{m_{\tilde{t}_2}^2}{m_{\tilde{t}_1}^2}\right)\right\}/m_t^2 \,],
\label{eq:higgsmass}
\eeqa
where $s_t (c_t)=\sin\theta_t (\cos\theta_t)$ and $y_t$ is the top Yukawa coupling. For tan$\beta$ sufficiently large that cos$^2(2\beta)\approx 1$, but not so large that (s)bottom loops are significant, the Higgs boson mass at one-loop is therefore determined by the three parameters $\mstl,\,\msth,$ and $\theta_t$.  

For degenerate stops, the logarithmic stop correction in the first term in the square parenthesis is  dominant. In this degenerate scenario, increasingly heavy stops masses can be made consistent with the measured Higgs mass $m_h=125$ GeV by appropriately decreasing the tree-level contribution to match the increasing loop contribution.  However, as the mass splitting between the two stops increases, the remaining two terms in the loop correction grow stronger. A key observation is that the final term switches sign and becomes negative for $m_{\tilde{t}_2}\,\gsim\, 2.7 m_{\tilde{t}_1}$. For non-vanishing stop mixing and $\msth\gg\mstl$, this negative term can dominate. 
Therefore, for nonzero mixing, there exists an $\textit{upper}$ limit on $\msth$ (as a function of $\stl$ and $\theta_t$), beyond which it is impossible to accommodate $m_h=125$ GeV in the MSSM.\footnote{This is simply an alternate formulation (in terms of the physical stop masses) of the more familiar statement that the Higgs mass cannot be raised arbitrarily by increasing the stop trilinear term $A_t$; beyond a certain value, further increasing $A_t$ lowers the Higgs mass. Note that this statement is only valid for nonzero mixing, and remains applicable at low tan $\beta$.} In other words, a measurement of $\mstl$ and some knowledge of $\theta_t$ allows for an upper limit on $\msth$. Ruling out $\msth$ in this window rules out the MSSM.  This statement is independent of the rest of the supersymmetric spectrum at one-loop (see related discussion below). In this case, one can conclude that the supersymmetric theory must include additional corrections to the Higgs mass, as, for instance, in the Next-to-Minimal Supersymmetric Standard Model (NMSSM). Thus, even partial information on the three parameters $\mstl,\,\msth,$ and $\theta_t$ can suffice to make meaningful statements about the underlying supersymmetric model. 

\begin{figure}[t]
\centering
\includegraphics[width=0.48\linewidth]{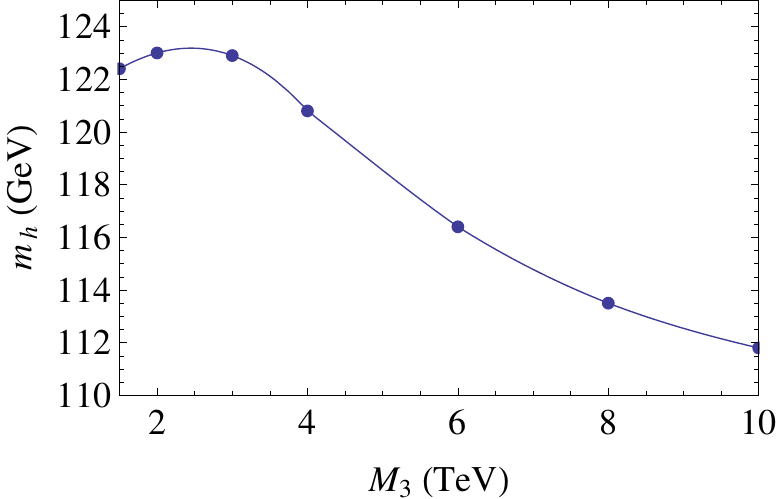}
\caption{Higgs mass dependence on the gluino mass parameter $M_3$ for $\msth=851$ GeV, $\mstl=459$ GeV, and sin\,$\theta_t=0.724$, as calculated with {\tt SUSY-HIT} \cite{Djouadi:2006bz}.}
\label{fig:mhmgluino}
\end{figure}

In this paper, we will make use of the analytic one-loop formula in Eq.\,(\ref{eq:higgsmass}) to calculate the Higgs mass. While a crude approximation, it is sufficient to illustrate our  ideas.   We also take the $120\leq m_h\leq 130$ GeV mass window as potentially compatible with the measured mass of the Higgs boson; we allow this perhaps surprisingly generous 10 GeV window to account for several corrections not captured by this simple formula, which are known to amount to a few GeV. For example, the Higgs mass is sensitive to both the uncertainty in and the running of the top Yukawa (we use $m_t=173$ GeV); these are known to affect the Higgs mass by a few GeV \cite{Vega:2015fna, Draper:2013oza, Lee:2015uza}. Likewise, at higher loop order the Higgs mass is sensitive to the gluino mass, particularly for large stop mixing. Assuming the gluino is not too heavy (remains $\lsim 4$ TeV), we find that the associated uncertainly in the Higgs mass remains a few GeV. We illustrate this dependence for a specific choice of stop masses and mixing angle in Fig.\,\ref{fig:mhmgluino} (see also Ref.\,\cite{Degrassi:2001yf} for a more detailed study; its results are in agreement with the above statement). For very heavy gluinos, $\mathcal{O}(10)$ TeV, the Higgs mass rises logarithmically with the gluino mass (see Ref.\,\cite{Degrassi:2001yf}), and it might be possible to recover the correct Higgs mass with such a spectrum.  However, it is difficult to motivate a scenario where the gauginos are much heavier than the squarks, and we do not pursue this direction further. Varying tan\,$\beta$ over the range $15\leq$ tan\,$\beta\leq 40$ does not change the Higgs mass by more than $\sim 1$ GeV.  The 10 GeV window of uncertainty is broad enough to encompass all of these factors.\footnote{We have verified with a scan with {\tt SUSY-HIT} that all compatible points fall within this 10 GeV window on the Higgs mass.}

The left-handed sbottom is inextricably linked with the stop sector as it is part of a doublet with the left-handed stop.  For  simplicity, we decouple the right-handed sbottom from our analysis. In this case, the lighter sbottom is purely left-handed; in terms of the parameters we are working with, its mass can then be written (for the case of vanishing sbottom mixing and cos$(2\beta)\approx -1$) as \cite{Batell:2015zla}
\beq
m_{\tilde{b}_1}^2=m_{\tilde{t}_2}^2\sin^2\theta_{\tilde{t}}+m_{\tilde{t}_1}^2\cos^2\theta_t-m_t^2+m_W^2.
\label{eq:sbottomrelation}
\eeq
Thus the lighter sbottom mass is fixed by the same three parameters that fix the Higgs mass, providing another constraint in the system. The upper limit discussed above for $\msth$ (imposed by the Higgs boson mass) can also be translated to an upper limit on $\msbot$.

In this paper, we work with the most minimal possible spectrum, decoupling all particles other than $\stl, \sth, \sbot,$ and a bino-like neutralino $\chi_0$, which we take to be the lightest supersymmetric particle (LSP).\footnote{Higgsinos are motivated to be light from naturalness considerations. The presence of both light charginos and neutralinos would lead to additional collider signatures.  In this paper, for simplicity, we decouple the Higgsinos and keep only a light bino to demonstrate that progress is possible even with this minimal scenario. For investigations of scenarios where Higgsinos are light and part of the phenomenology, see \eg\,\cite{Kobakhidze:2015scd,Goncalves:2016tft,Han:2016xet,Goncalves:2016nil}. We also remain agnostic about whether the LSP can account for some or all of dark matter.}

\subsection{Indirect Constraints on Light Stops}

To appreciate the range of possible LHC signals, it is useful to first discuss indirect constraints on light third generation squarks. In particular, when stop mixing is significant, as might be suggested by the Higgs boson mass if the stops are light, there can be significant contributions to the $\rho$ parameter or a modification of the Higgs boson production rate.

\begin{figure}[t]
\centering
\includegraphics[width=0.5\linewidth]{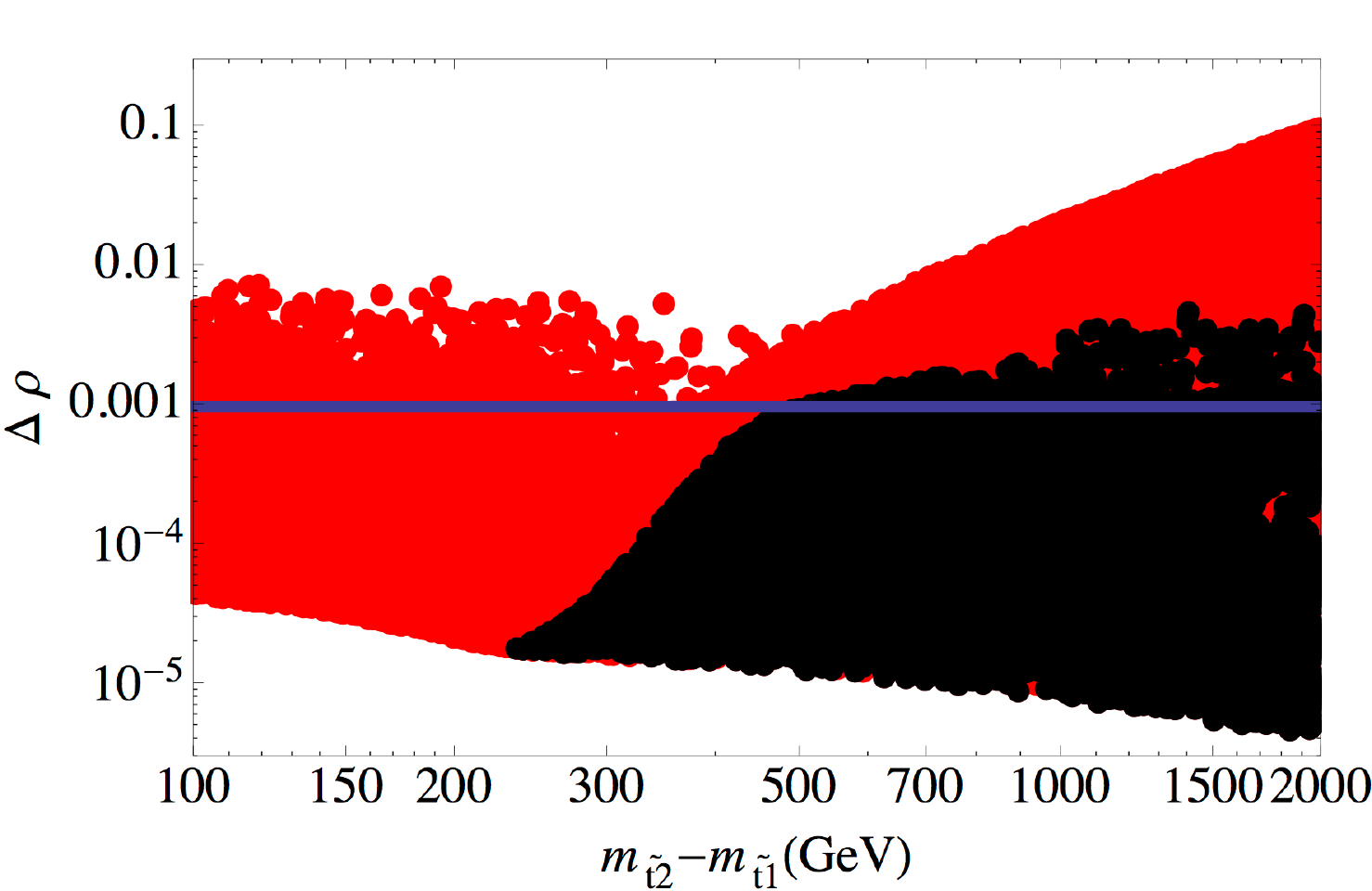}
\caption{$\Delta\rho$ for all scanned points (red) and those compatible with $120\,\textless\, m_h\,\textless\, 130$ GeV (black), as a function of the mass splitting between the two stop mass eigenstates. All points have $\mstl\,\textless \,1$ TeV, and we have scanned over all possible values of the mixing angle, $0\leq s_t^2\leq 1$. The horizontal blue line denotes $\Delta\rho\,=\,9.6\times 10^{-4}$, the $2\sigma$ upper limit from measurements.}
\label{fig:deltarho}
\end{figure}

The one-loop contribution to the $\rho$ parameter is \cite{Barbieri:1983wy,Lim:1983re,Drees:1990dx,Heinemeyer:2004gx}
\beq
\Delta\rho=\frac{3\,G_F}{8\sqrt{2}\pi^2}\left[-s_t^2\,c_t^2\,F_0(m_{\tilde{t}_1}^2,m_{\tilde{t}_2}^2)+c_t^2\,F_0(m_{\tilde{t}_1}^2,m_{\tilde{b}_1}^2)+s_t^2\,F_0(m_{\tilde{t}_2}^2,m_{\tilde{b}_1}^2)\right]\,,
\eeq
where
\beq
F_0(x,y)=x+y-\frac{2 x y}{x-y}\,\log\frac{x}{y}\,.
\eeq
We take the constraint from Ref.\,\cite{Barger:2012hr}, 
$\Delta\rho = (4.2\pm2.7)\times 10^{-4}.$
We demand consistency with this number to $2\sigma$. In general, $\Delta\rho$ can increase for larger mass splitting or mixing angle. However, as discussed earlier, a large mass splitting with large mixing suppresses the Higgs boson mass on account of the large negative term in the loop contribution. Indeed, for $\mstl\,\textless\,1$ TeV, we find that points with the correct Higgs mass are correlated with small values of $\Delta\rho$. These features are shown in Fig.\,\ref{fig:deltarho}. 

The existence of a light stop can also modify the Higgs-gluon-gluon coupling, which is constrained to be somewhat close to its Standard Model (SM) value. The stop contribution to this coupling is \cite{Djouadi:2005gi,Djouadi:2005gj}
\beq
r_{gg}\equiv\kappa_g^2\equiv\frac{\Gamma(h\to gg)}{~~\Gamma(h\to gg)_{SM}}=\left(1+\frac{1}{A_{hgg(SM)}}\left[\frac{\lambda_{ht_1t_1}A_0[\mstl]}{\mstl\!^2}+\frac{\lambda_{ht_2t_2}A_0[\msth]}{\msth\!^2}\right]\right)^2,
\label{eq:rgg}
\eeq
\bea
\lambda_{ht_1t_1}&=&M_z^2 \cos 2 \beta \left(\frac{1}{2} c_t^2 - 
    \frac{2}{3} \sin^2\theta_W\, \cos 2 \theta_t\right) + m_t^2 - 
 s_t^2 c_t^2(\msth\!^2-\mstl\!^2)\,,\nonumber\\
\lambda_{ht_2t_2}&=&M_z^2 \cos 2 \beta \left(\frac{1}{2} s_t^2 + 
    \frac{2}{3} \sin^2\theta_W\, \cos 2 \theta_t\right) + m_t^2 + 
 s_t^2 c_t^2(\msth\!^2-\mstl\!^2)\,,\nonumber\\
 A_0[m]&=&-[\tau-f(\tau)]\tau^{-2},~~~f[\tau]=\text{arcsin}^2\sqrt{\tau},~~~\tau=\frac{m_h}{4\,m^2},~~~A_{hgg(SM)}\approx 1.38.
 \label{eq:httcouplings}
\eea
We include the contributions from both stop mass eigenstates, though the contribution from the lighter eigenstate tends to dominate due to the $1/m^2$ factor (the sbottom contribution, even when it is as light as $\stl$, is generally negligible). LHC data constrain $r_{gg}$ to within $\sim25\%$ of the SM value \cite{Khachatryan:2016vau}. The LHC is expected to probe this quantity to within $12-16\%$ ($6-10\%$) of the SM value with 300 fb$^{-1}$ (3000 fb $^{-1}$) of data, whereas the ILC and TLEP can probe it to percent level precision \cite{Dawson:2013bba}. Such constraints on $r_{gg}$ can therefore result in strong bounds on the stop mixing angle as a function of the two stop masses. 

\begin{figure}[t]
\centering
\includegraphics[width=0.33\linewidth]{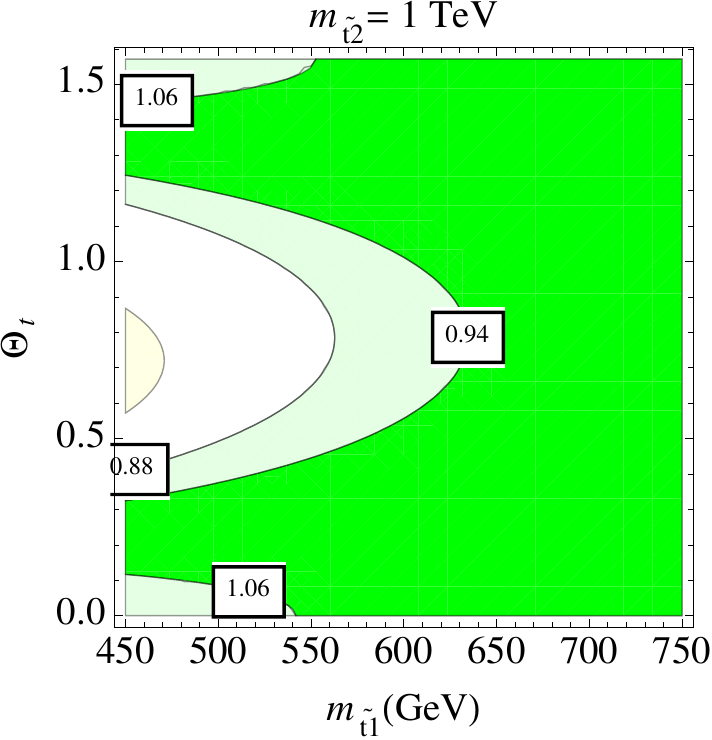}~\includegraphics[width=0.33\linewidth]{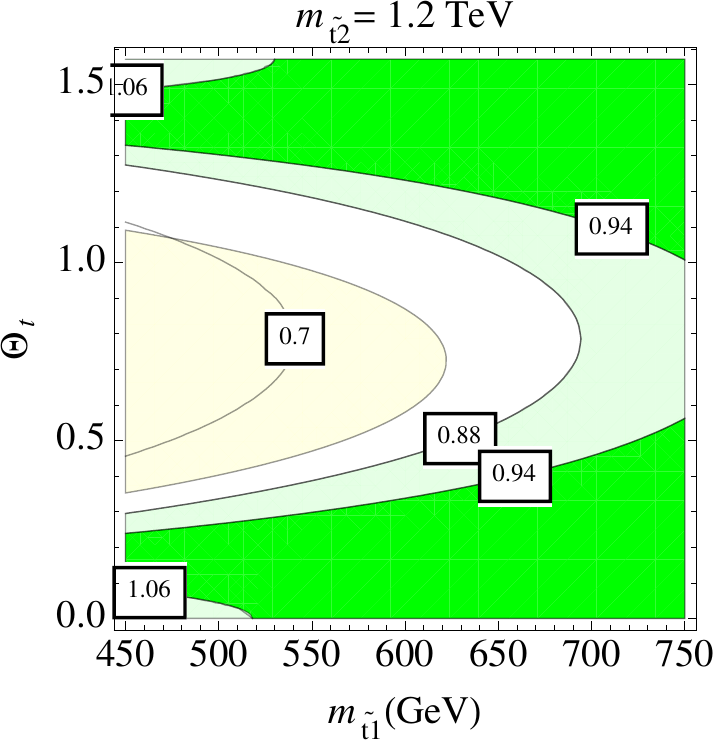}~\includegraphics[width=0.33\linewidth]{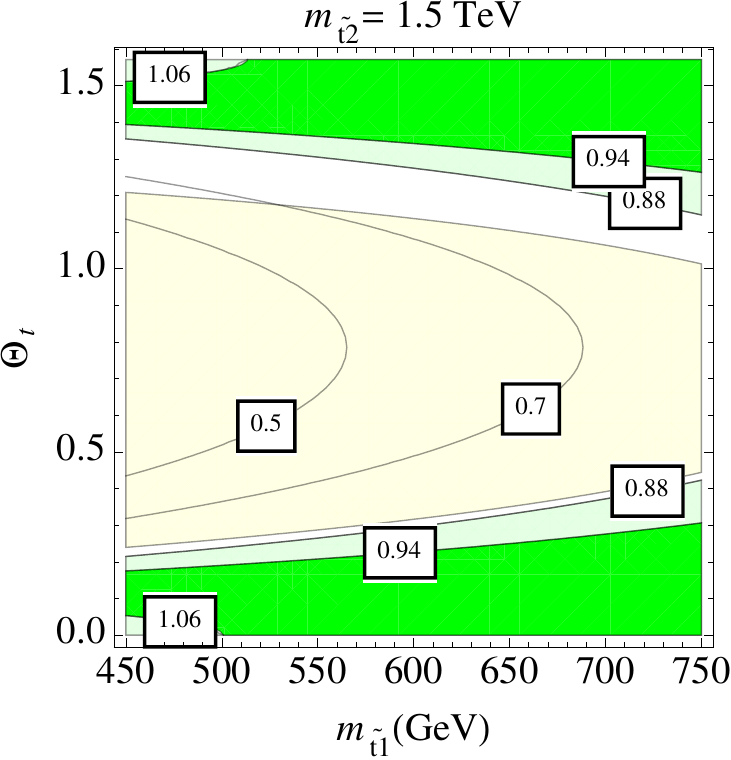}
\caption{Contours of $r_{gg}$ as a function of $\mstl$ and the stop mixing angle $\theta_t$, for three choices of $\msth:\, 1.0,\,1.2,\,$ and $1.5$ TeV in the left, center, and right panels. Light (dark) green regions denote $0.88\leq r_{gg}\leq 1.12$ ($0.94\leq r_{gg}\leq 1.06$), the optimistic reach with 300 (3000) fb$^{-1}$ data at the LHC. Yellow regions denote parameter space incompatible with $\Delta\rho$ constraints.}
\label{fig:hgg}
\end{figure}

We illustrate the potential power of such constraints in Fig.\,\ref{fig:hgg}; the light (dark) green shades denote regions that would be compatible with future LHC runs with 300 (3000) fb$^{-1}$ data. Note that the corresponding constraint on the mixing angle becomes stronger as $\stl$ becomes lighter or $\sth$ becomes heavier. Notably, we see that the allowed regions of parameter space can cleanly separate into two distinct bands corresponding to small and large mixing angles, $\ie$ a mostly left-handed or right-handed $\stl$. We also show regions incompatible with $\Delta\rho$ constraints in yellow, which become stronger as the mass splitting between the stop mass eigenstates increases, as seen earlier in Fig.\,\ref{fig:deltarho}. 

 In the next three sections, we demonstrate how the above ideas can be implemented in three distinct scenarios at the LHC, corresponding to qualitatively very different signals from $\stl,\sbot$, and $\sth$. In all cases, we demand compatibility with both $\Delta\,\rho$ and $r_{gg}$.

%%%%%%%%%%%%%%%%%%%%%%%%%%%%%%%%
\section{Implications of a Light Stop Signal}
\label{sec:stop1}
%%%%%%%%%%%%%%%%%%%%%%%%%%%%%%%%

The latest LHC results impose increasingly stringent constraints on light stops: limits exist for $\mstl\sim m_t+m_\chi$ \cite{Aaboud:2017ayj,ATLAS-CONF-2017-037}, 3-body decay into $bW\chi_0$ \cite{Aaboud:2017ayj,Aaboud:2017nfd}, 4-body decay into $bff'\chi_0$ \cite{Aaboud:2017nfd}, as well as flavor violating decays into $c\chi_0$ \cite{Sirunyan:2017kiw}. Together, these bounds essentially rule out stop masses below $\mstl\lsim 450$ GeV. In this section, we therefore focus on the mass window $450\leq \mstl\leq 600$ GeV, where a light stop is potentially compatible with existing constraints, and discuss the implications of its discovery at the LHC. As described in the previous section, the existence of a light stop invites non-trivial constraints from $r_{gg}$. For this section, we assume the optimistic reach with 3000 fb$^{-1}$ of data at the LHC as reported in Ref.\,\cite{Dawson:2013bba}, which will constrain $0.94\leq r_{gg}\leq 1.06$.

\begin{figure}[t]
\centering
\includegraphics[width=0.5\linewidth]{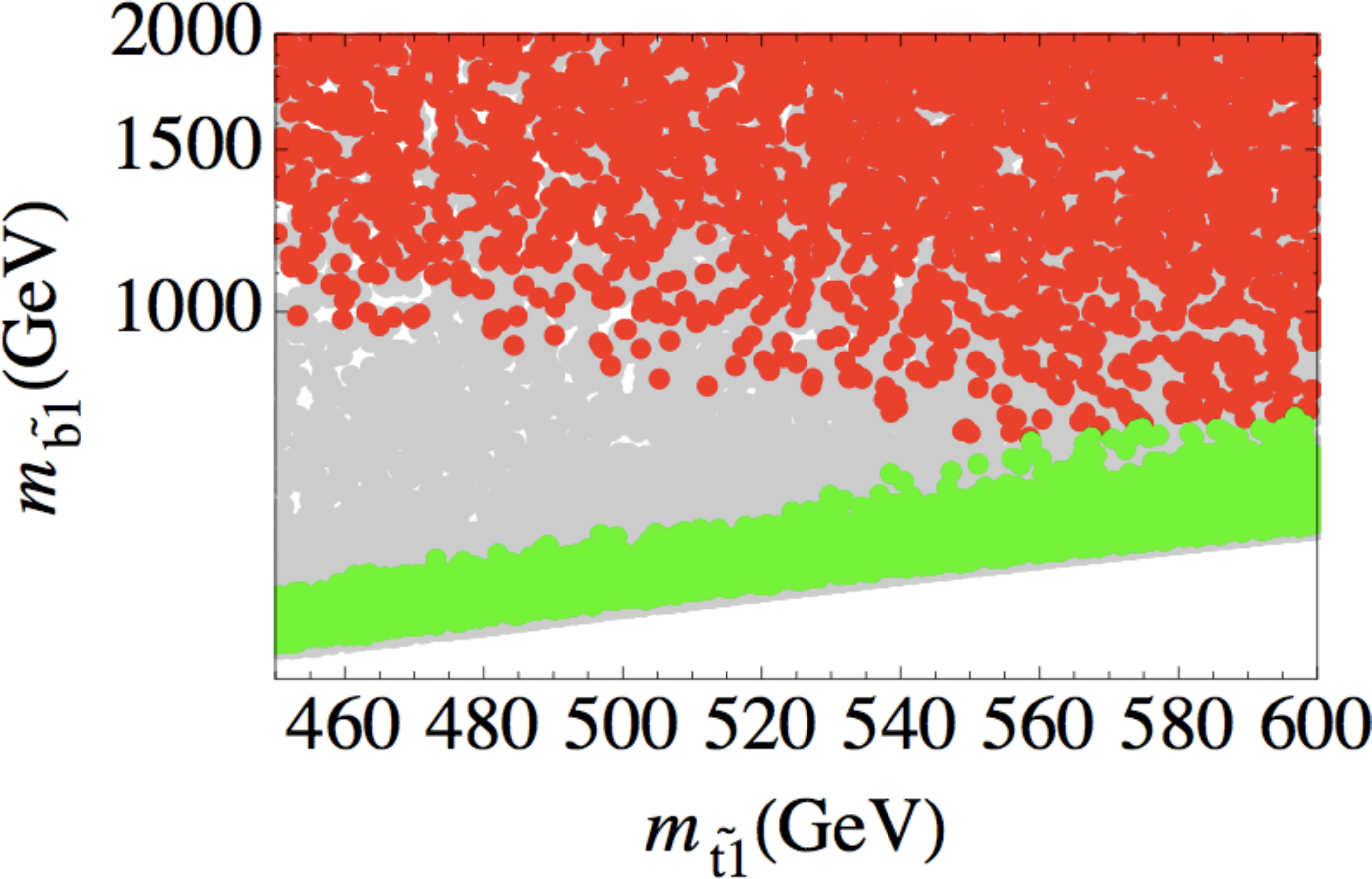}~~~~\includegraphics[width=0.5\linewidth]{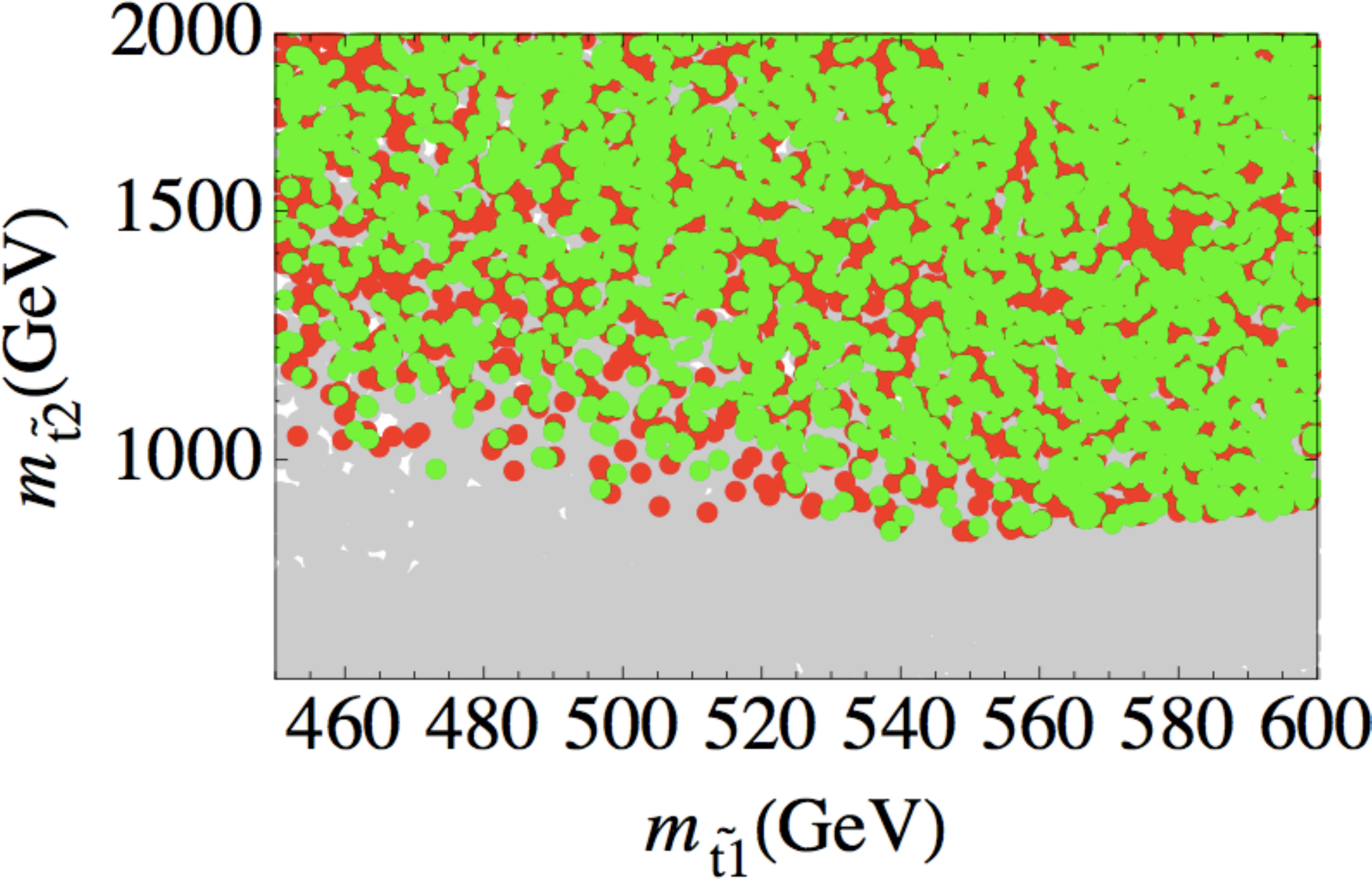}\\
\vskip0.1in
\includegraphics[width=0.5\linewidth]{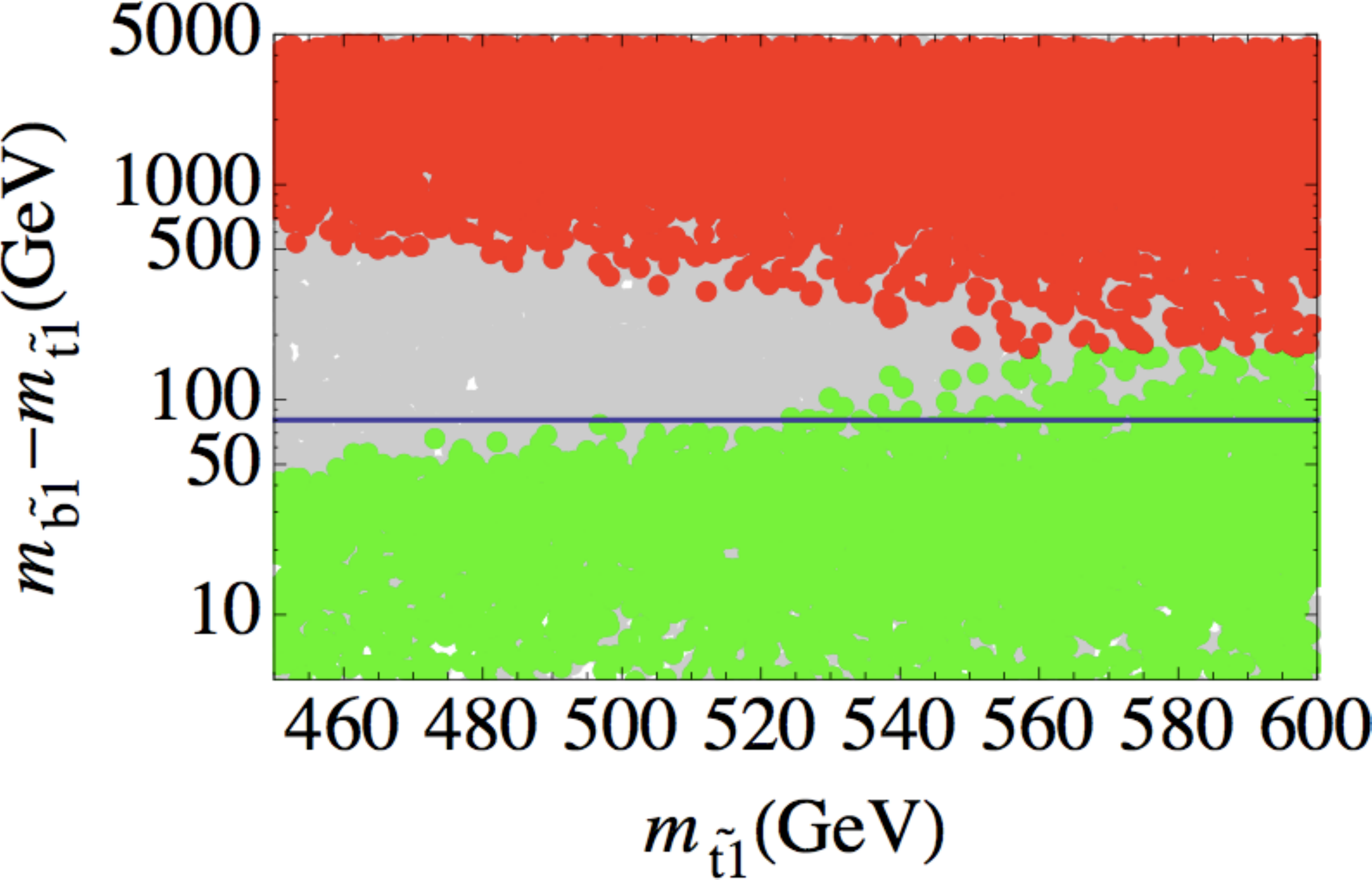}~~~~\includegraphics[width=0.5\linewidth]{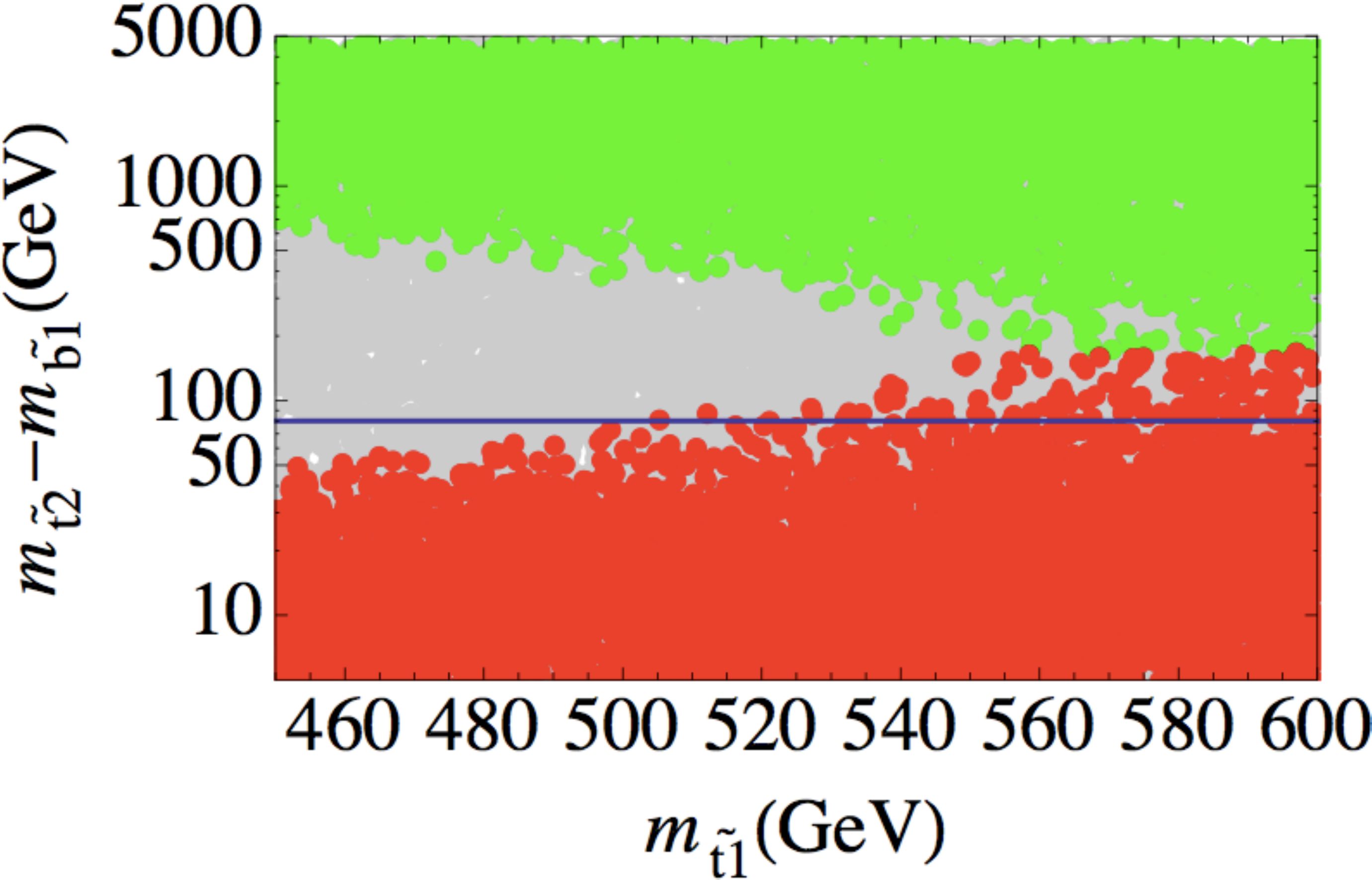}
\caption{Masses of sbottom (top left) and heavier stop (top right), and mass splittings between $\sbot$ and $\stl$ (bottom left) and between $\sth$ and $\sbot$ (bottom right) for light $\stl$ masses. All points are compatible with $\lambda_{hgg}$ and $\Delta\rho$ constraints. Green (red) points are compatible with $120\leq m_h\leq 130$ GeV for mostly left-handed (right-handed) $\stl$; grey points have $m_h$ outside this window. The horizontal line in the bottom row corresponds to mass splitting equal to the $W$ boson mass, which has important implications for collider searches.}
\label{fig:sbstoponium}
\end{figure}

The constraint on $r_{gg}$ can be mapped onto the physical parameters $\msbot$ and $\msth$. This is plotted in the top row of Fig.\,\ref{fig:sbstoponium}, obtained by performing a scan over parameter space, demanding consistency with both $0.94\leq r_{gg}\leq 1.06$ and $\Delta\rho\,\textless\, 9.6\times 10^{-4}$. We also find it  instructive to look at the mass splittings $\Delta m_{\sbot\stl}\equiv\msbot-\mstl$ and $\Delta m_{\sth\sbot}\equiv\msth-\msbot$, which are plotted in the bottom row; the horizontal blue line denotes mass splitting equal to $m_W$. Points that satisfy $120\leq m_h\leq 130$ GeV in the MSSM are plotted in green (red) for primarily left-handed (right-handed) $\stl$, while points outside this mass window are shown in gray. We see that while arbitrary $\msbot$ and $\msth$ can be realized for the mixing angles allowed by $r_{gg}$ and $\Delta\rho$, interesting patterns emerge with the additional requirement of reproducing the Higgs mass. As discussed in the previous section, this imposes an upper limit on $\msth$ and $\msbot$. It is convenient to separate the discussion into cases where the light stop $\stl$ is mostly left-handed ($\theta_t\leq \pi/4$, green points) or mostly right-handed ($\theta_t\geq \pi/4$, red points).

For mostly left-handed\footnote{While additional observations are required to determine whether a stop is left- or right-handed, theoretical considerations may prefer one over the other -- for instance, in gauge mediation, the lighter stop is generally left-handed \cite{Knapen:2016exe}.} $\stl$, this limit is not very meaningful for $\msth$, which can be at several TeV  (the constraints do impose a {\it lower} bound on $\msth$). However, it is sharp for $\msbot$, constraining $\Delta m_{\sbot\stl}\,\lsim 200$ GeV in the $\stl$ mass window of interest, as seen in the left panels of Fig.\,\ref{fig:sbstoponium}. In addition to revealing the existence of a light sbottom, these correlations also reveal information about its decay channels: below (above) the $m_W$ line, $\sbot$ decays primarily to $b\chi_0$ ($\stl W$).\footnote{We also see points with $\msbot\textless\,\mstl$; we do not address them further in this paper.} For $\mstl\lsim 500$ GeV, the relevant splitting is constrained to be smaller than $m_W$, and the sbottom decays as $\sbot\to b \chi_0$. This decay is strongly constrained by the LHC, with the latest bounds \cite{CMS-PAS-SUS-16-032} ruling out  $\msbot\lsim 1$ TeV for $m_{\chi_0}\lsim 500$ GeV, effectively eliminating this region ($\mstl\lsim 500$ GeV) of MSSM parameter space. 
For $\mstl\gsim 500$ GeV, the sbottom can decay primarily as $\sbot\to\stl W$ if $\Delta m_{\sbot\stl}\,\textgreater\,m_W$, which requires the stops to be split due to large mixing. As argued in the previous section, such large mixings, in turn, enforce strong upper limits on $\msth$ for compatibility with the MSSM Higgs mass. We find that the $\Delta m_{\sbot\stl}\textgreater\, m_W$ region is correlated with $\msth\lsim 1.2$ TeV. Such masses are potentially within reach of the 14 TeV LHC, although discovery will be challenging and will require dedicated searches. For a discussion of possible detection strategies in various scenarios, see $\eg$ Ref.\,\cite{Cheng:2016npb}. 

For a mostly right-handed $\stl$, both the $\sbot$ and $\msth$ are heavy ( $\gsim 700$ GeV; top panels, red points), and searching for their signals is challenging. In this case, the absence of such signals at the LHC does not lead to any meaningful conclusions about the MSSM. In contrast, it is the $\it{lower}$ bounds on these masses that are relevant.  Should a  $\sbot$ or $\sth$ be discovered with mass lighter than what is shown in the figure, this would be inconsistent with the MSSM, pointing to physics -- and contributions to the Higgs mass -- beyond the MSSM.\footnote{A light right-handed sbottom could accidentally exist in the spectrum. Furthermore, we have not considered sizable mixing in the sbottom sector, which could lead to lower sbottom masses than shown in the figure via mixing with the heavier sbottom. In this case, there may also be corresponding large contributions to $r_{gg}$ if tan$\beta$ is large, which must be taken into account.}  For a mostly right-handed $\stl$, the sbottom mass is more closely aligned with the heavier stop mass, and we find $\Delta m_{\sth\sbot}\lsim 200$ GeV (bottom right panel, red points). Moreover, for $\mstl\lsim 500$ GeV, $\Delta m_{\sth\sbot}\,\textless\, m_W$, which implies $\sth\to\sbot W$ is not allowed in this window, motivating $\sth$ searches in the $\stl Z$ (and possibly $\stl h$) channels (we will explore such signals in Sec.\,\ref{sec:stop2}). Likewise, increasing $\Delta m_{\sth\sbot}\textgreater\, m_W$, which appears possible for $\mstl\gsim 500$ GeV, again requires mixing in the stop sector, resulting in $\msth\lsim 1.2$ TeV for compatibility with the Higgs mass in the MSSM, which represents an attractive target for the LHC. In Section\,\ref{sec:sbottom}, we perform a detailed study of a scenario where both $\sbot\to\stl W$ and $\sth\to\sbot W$ are open, leading to a multitude of leptonic signals at the LHC.

With the above considerations in mind, we divide our discussion of the interpretations of a $\stl$ signal into two distinct mass windows.

\vskip0.1in
\noindent\textit{A. $\mstl\sim450-500$ GeV}:
\vskip0.05in

In this window, current data allow for the flavor violating decay $\stl \rightarrow c\chi_0$. 

\begin{itemize}
\item In the MSSM, if $\stl$ is mostly left-handed, this discovery implies a light sbottom ($\msbot\lsim 580$ GeV) that decays as $\sbot\to b\chi_0$, which is already ruled out by the latest LHC bounds \cite{CMS-PAS-SUS-16-032,CMS-PAS-SUS-16-016,CMS-PAS-SUS-16-015,CMS-PAS-SUS-16-014}. Discovering such a light stop would therefore imply that either the stop is mostly right-handed (red points) or the underlying theory is not the MSSM (gray points with $\Delta m_{\sbot\stl}\,\textgreater\, m_W$, but with the wrong Higgs boson mass). 
\item If $\stl$ is mostly right-handed, $\Delta m_{\sth\sbot}\,\textless\, m_W$ in the MSSM, thus this tells us that $\sth$ decays primarily to $\stl Z, \stl h$.  However,  both $\sth$ and $\sbot$ could be extremely heavy (several TeV) and escape detection. 
\end{itemize}

\vskip0.1in
\noindent\textit{B. $\mstl\sim500-600$ GeV}:
\vskip0.05in

This heavier stop mass window follows a similar pattern as above, but with the crucial difference  that $\sbot\to\stl W$ (for left-handed $\stl$) and $\sth\to\sbot W$ (for right-handed $\stl$) can both be kinematically open. These channels can be difficult to probe if the involved mass splittings are not very large.  A $\stl$ discovery in this mass window carries the following implications:

\begin{itemize}
\item For a mostly left-handed $\stl$, there are two possibilities for the sbottom in the MSSM. If $\sbot\to b\chi_0$, the corresponding $bb+\met$ signal, if not observed or ruled out already, should easily be visible at the LHC (see \cite{CMS-PAS-SUS-16-032} for the latest constraints on this channel). If instead $\sbot\to\stl W$, sbottom discovery may be difficult; however, as discussed earlier, $\msth\lsim 1.2$ TeV and might have better detection prospects. 
\item  For a mostly right-handed $\stl$, we can conclude that either $\sth\to\sbot W$ is not allowed (in this case both $\sth$ and $\sbot$ can be very heavy), or $\msth\lsim 1.2$ TeV. Note that in this mass window the lower limits on the allowed $\sbot$ and $\sth$ masses are also relaxed relative to the lighter stop mass windows (top right panel).  
\end{itemize}

%%%%%%%%%%%%%%%%%%%%%%%%%%%%%%%%
\section{Using Sbottoms to Help Reconstruct the Stop Sector}
\label{sec:sbottom}
%%%%%%%%%%%%%%%%%%%%%%%%%%%%%%%%

In this section, we discuss how a left-handed sbottom might be observed at the LHC, and how this observation, in conjunction with searches for a heavy stop, can test the MSSM. This strategy can provide successful tests of the MSSM even in the absence of detailed direct information about the lighter stop. 

For a predominantly left-handed sbottom, the decay $\sbot\to\stl W$ dominates if kinematically accessible and $\stl$ has non-negligible left-handed content. For this section, we have in mind a spectrum where $\sbot\to\stl W$ and $\stl\to t\chi_0$, but discovery of direct light stops production is elusive because the spectrum is squeezed. The presence of $W$s in the sbottom decay chain suggests that multilepton channels might be a fruitful way to search for such sbottoms. We adopt two search strategies involving leptons as applied at CMS: same-sign dileptons \cite{Khachatryan:2016kod} (recently updated in \cite{CMS-PAS-SUS-16-020}), and a multilepton search strategy \cite{CMS-PAS-SUS-16-003} (recently updated in \cite{CMS-PAS-SUS-16-022}), which searches for an excess in $\geq 3\,l$+jets+$\missET$ at the 13 TeV LHC, and explore whether a sbottom signal can be uncovered with these approaches in the future.\,\footnote{CMS analyses do not explicitly interpret their searches using the topology we have in mind ($\sbot\to\stl W,\,\stl\to t\chi_{0}$), but Ref.\,\cite{CMS-PAS-SUS-16-003}  casts its results in terms of the similar ( $\tilde{b}\rightarrow t\,\chi^-,\,\chi^-\rightarrow W\,\chi_0$), placing the limit $\msbot\lsim 550$ GeV using 2.3/fb of 13 TeV LHC data. By simulating events for representative benchmark points for the two topologies, we have verified they give similar lepton spectra and $\missET$ distributions and similar efficiencies (within $\sim10\%$), for $\mstl\approx m_{\chi^-}$. }

Before exploring the reach of these search strategies, we first discuss the implications of observing such a signal. A multilepton excess interpreted as a $\sbot\to\stl W,\,\stl\to t\chi_0$ signal (further corroboration, such as an independent measurements of $\stl$ and the presence of $b$ tags in the excess, will help solidify this interpretation) implies that both $\stl$ and $\sbot$ are somewhat left handed (necessary for $\sbot\to\stl W$ decays), and $\Delta m_{\tilde{b}_{1} \tilde{t}_{1}}$ must be sufficiently large (for the leptons to be hard enough to be observed). Taken together, these imply appreciable mixing in the stop sector, as a purely right-handed $\stl$ precludes this decay channel altogether, while a purely left-handed $\stl$ does not result in a sufficiently large mass splitting with $\sbot$.

\begin{figure}[t]
\centering
\includegraphics[width=0.6\linewidth]{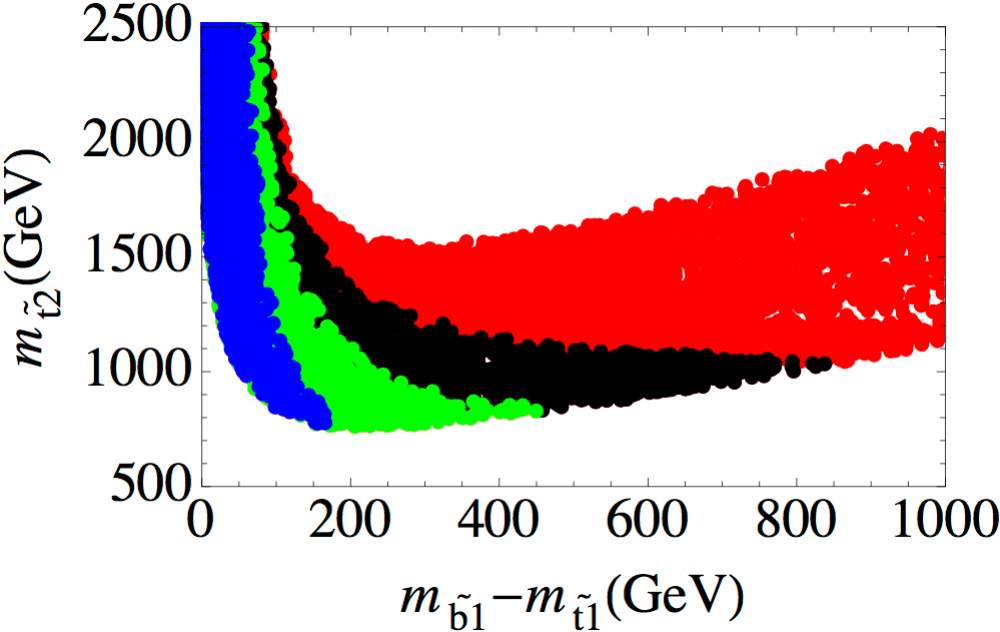}
\caption{$\msth$ as a function of the mass splitting $\msbot-\mstl$  for points with $120\,\textless\, m_h\,\textless\,130$\,GeV  and $\mstl\,\textless\,1$\,TeV in the MSSM. Red, black, green, and blue regions correspond to sbottom masses $\msbot\textgreater\,1000,\,750\,\textless\,\msbot\,\textless\,1000,\,500\,\textless\,\msbot\,\textless\,750,$ and $\msbot\,\textless\,500$ GeV respectively.}
\label{fig:stop2splitting}
\end{figure}

\begin{figure}[t]
\centering
\includegraphics[width=0.58\linewidth]{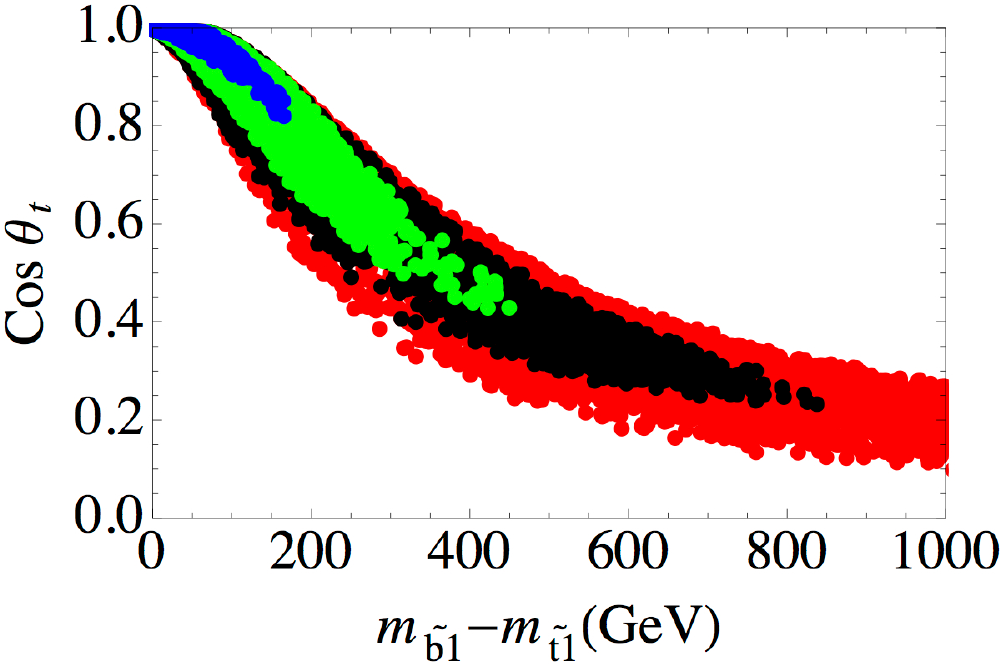}
\caption{Cosine of stop mixing angle as a function of the mass splitting $\msbot-\mstl$ for points with $120\,\textless\, m_h\, \textless\, 130$ GeV and $\mstl\,\textless\,1$\,TeV in the MSSM. Color coding is the same as in Fig.\,\ref{fig:stop2splitting}.}
\label{fig:mixingsplitting}
\end{figure}

In the MSSM, these observations have consequences for the second (heavier) stop. We find $\msth$ is correlated with the $\Delta m_{\sbot\stl}$ mass splitting, as shown in Fig.\,\ref{fig:stop2splitting} (for $\mstl\,\textless\,1$ TeV and $120\leq m_h\leq 130$ GeV in the MSSM as calculated from the 1-loop formula in Eq.\,(\ref{eq:higgsmass})). The colored regions (red, black, green, and blue) correspond to different sbottom masses ($\msbot\textgreater\,1000,\,750\,\textless\,\msbot\,\textless\,1000,\,500\,\textless\,\msbot\,\textless\,750,$ and $\msbot\,\textless\,500$ GeV respectively). At small $\Delta m_{\sbot\stl}$ ( $\lsim 150$ GeV --  more difficult to probe via multilepton searches), stop mixing is small, $\stl$ is mostly left-handed, and the desired Higgs mass can be obtained with a heavy (several TeV) $\sth$. The splitting $\Delta m_{\sbot\stl}$ can be made larger by increasing the stop mixing angle (this correlation is plotted in Fig.\,\ref{fig:mixingsplitting}); in this case, as discussed in Section\,\ref{sec:framework}, consistency with the Higgs mass enforces an upper limit on $\msth$, which is indeed clearly visible in Fig.\,\ref{fig:stop2splitting}. Alternatively, the splitting can be raised without significant stop mixing by making $\stl$ mostly right-handed; however, in this case, $\sth$ becomes approximately degenerate with $\sbot$ and thus again faces an upper mass limit. For large $\Delta m_{\sbot\stl}$, there is therefore an upper limit on $\msth$, which grows stronger for lighter $\sbot$, as seen in the various colored bands in Fig.\,\ref{fig:stop2splitting}. This becomes particularly sharp for a sub-TeV sbottom, which forces $\msth$ into a narrow wedge-shaped region -- for instance, for $\msbot\textless\,750$ GeV and $\Delta m_{\sbot\stl}\gsim 150\,(200)$ GeV, Fig.\,\ref{fig:stop2splitting} tells us that $\msth\lsim 1.3\, (1.1)$ TeV.   

Thus, the observation of a multilepton+$\missET$ signal associated with a sub-TeV sbottom and large $\Delta m_{\sbot\stl}$ leads to a robust upper limit on $\msth$ in the MSSM.  These stops  may well be within reach of the LHC. Detailed analysis of the  multilepton excess can shed further light on the properties of $\msth$: inferring $\Delta m_{\sbot\stl}$ (from, e.g., the lepton $p_T$ distribution) and the sbottom mass (from, e.g., the signal rate) not only narrows the allowed range of $\msth$ (Fig.\,\ref{fig:stop2splitting}) but also constrains the stop mixing angle (Fig.\,\ref{fig:mixingsplitting}). It is therefore possible to not only predict a relatively narrow mass window for $\sth$, but also get a profile of its decay channels.  Ruling out such a $\sth$ is sufficient to rule out the MSSM.

A TeV scale $\sth$ can be probed in several ways. If it decays primarily via $\tilde{t}_2\rightarrow \tilde{t}_1 Z$, leptonic decays of boosted $Z$-bosons offer a promising search strategy \cite{Perelstein:2007nx}. For $\tilde{t}_2\rightarrow \tilde{b}_1 W$, the cascade $\tilde{t}_2\rightarrow \tilde{b}_1 W$, $\tilde{b}_1\rightarrow \tilde{t}_1 W$, $\tilde{t}_1\rightarrow t \chi_0$ can produce several high $p_T$ leptons; in this case, there could be excesses in both $\geq\,3$ and $\geq 4$ lepton searches. Likewise, $\tilde{t}_2\rightarrow \tilde{t}_1 h$ can be probed by reconstructing the $h$ with boosted $b$-jets (see e.g. \cite{Ghosh:2013qga}).  Note that even when $\tilde{t}_2\rightarrow \tilde{t}_1 h$ is large, $\sth$ maintains the decay $\tilde{t}_2\rightarrow \tilde{t}_1 Z$, so the search for leptonic $Z$, boosted hadronic $Z$s, or a mix of $Z$ and $h$ might be possible.
 
As an aside,  it is important to verify that the original multilepton excess originates primarily from $\sbot$ decays rather than from $\sth$ decays. It is possible to be initially fooled:   boosted leptons from $\sth$ decays in the blue region of Fig.\,\ref{fig:stop2splitting} might mimic a $\sbot \rightarrow \stl W$ signal, leading to the erroneous interpretation that $\Delta m_{\sbot\stl}$ is large. However, in such cases, the $\sth$ search strategies discussed in the previous paragraph can reveal the heavier stop, and help clarify the extent to which it might give rise to a multilepton excess.

\subsection{Benchmark Case Study}

We choose an MSSM benchmark that produces the correct Higgs boson mass and has third generation squarks below the TeV scale. The masses and branching ratios of the relevant particles, generated with {\tt SUSY-HIT} \cite{Djouadi:2006bz}, are listed in Table\,\ref{tab:benchmark}. For this spectrum, we simulated 13 TeV LHC collision events with the {\tt Madgraph5$\_$aMC$@$NLO} package \cite{Alwall:2014hca}, with detector simulation using {\tt Delphes-3.2.0} \cite{deFavereau:2013fsa}.\,\footnote{We change the b-tagging efficiency to $0.7$ to match the CMS analysis we rely on for cuts and background, but otherwise use default parameters from our implementation of {\tt Madgraph5} and {\tt Delphes}. To ensure that the default implementations are similar to the CMS analysis that we mirror, we simulate the $ttZ/h$ background, which is one of the dominant background for our analyses, and verify that the efficiency for this background contribution matches that from the CMS analysis.} 

\begin{table}[t]
\begin{center}
\begin{tabular}{|l|} \hline
~~~$\sth:~~~\msth=1022.2$ GeV,~~~ BR:~~$\stl Z$ 79\%,~$\sbot W$ 15\%,~$\stl h$ 2\%, ~~ sin\,$\theta_t=0.75$~~~~~\\
~~~$\sbot:~~~\msbot=885.4$ GeV,~~~ BR:~~$\stl W$ 99.5\% \\
~~~$\stl:~~~\mstl=646.1$ GeV,~~~ BR:~~$t \chi_0$ 100\% \\
~~~$\chi_0:~~~m_\chi=445.7$ GeV,~~~LSP\\
~~~$h:~~~m_h=123.2$ GeV\\
\hline
\end{tabular}
\caption{Benchmark point: Mass spectrum and branching ratios. For this point, we set tan$\,\beta=15$, $\mu=800$ GeV, and all other masses (including the gluino mass) to 2 TeV.}
\label{tab:benchmark}
\end{center}
\end{table}

\vskip0.3cm
\noindent\textit{Same-sign dileptons search}
\vskip0.3cm

We first investigate the same-sign dilepton (SSDL) search as discussed in the CMS paper \cite{Khachatryan:2016kod} (recently updated in \cite{CMS-PAS-SUS-16-020}), also advocated by recent phenomenological studies \cite{Guo:2013iij,Cheng:2016npb} as a promising search strategy for heavier superpartners. We begin by mirroring the analysis in this CMS paper, imposing the following cuts:
\begin{itemize}
\item Require same sign dileptons with $p_T \,\geq \, 25$ GeV. 
\item Impose a $Z$ veto: reject events with opposite sign, same-flavor dilepton pairs with an invariant mass between 76 and 106 GeV.
\item Require two or more $b$-jets, $N_{b-jets}\geq 2$.
\item Constrain the missing transverse energy to $200\,\leq\, \missET\,\leq\,300$ GeV.
\item Require $300\,\leq\,H_T\,\leq\,1125$ GeV, where $H_T$ denotes the sum of transverse momenta of all the jets in the event.
\item Require 5 or more jets in the event, $N_{jets}\geq 5$.
\end{itemize}

The resulting number of signal and background events with 3000 fb$^{-1}$ of data at the 13 TeV LHC are shown in Table \ref{tab:resultsssdl}.  The signal primarily results from  $\sbot$ decays. %, with $\sth$ decays contributing less than $15\%$ of the signal. 
The expected background is taken from the CMS analysis \cite{Khachatryan:2016kod}, scaled up to 3000 fb$^{-1}$ of data. We also list the significance of the signal, calculated as $\mathcal{S}\equiv S\,/\,\sqrt{B+\sigma_{bg}^2 B^2}$; the second term in the denominator denotes systematic uncertainties, which are currently around $30\%$ \cite{CMS-PAS-SUS-16-003}, but should improve with future studies and additional data. We calculate the significance for $\sigma_{bg}=0$, 0.1, and 0.3 to span the range of possibilities. Our results show that close to a $3\sigma$ signal is possible with an improvement to $\sigma_{bg}=0.1$, while further improvements would push the significance towards a $5\sigma$ discovery. 

\begin{table}[t]
\begin{center}
\begin{tabular}{|l||c|c|c|c|c|c|c|} \hline
~~Search ~ & ~$\sigma_{\text{prod}}$/fb~ & efficiency($\epsilon$)& no. of signal & background & ~~$\mathcal{S}$~~&~~$\mathcal{S}$~~&~~$\mathcal{S}$~~\\
~ & ~ & ~($\times 10^{-4}$ ) & ~events (S)~ & ~events (B)~ & $\sigma_{bg}=0$ & $\sigma_{bg}=0.1$ & $\sigma_{bg}=0.3$ \\
\hline\hline
~SSDL~ & - & - & 72 & 235 & 4.7 & 2.6 & 1.0\\
\hline
~$\sbot$ contribution & 14 & 14 & 61 & - & - & -& -\\
~$\sth$ contribution & 5 & 8 & 11 & - & -& - & - \\
\hline
\end{tabular}
\caption{Same-sign dilepton analysis: Efficiency of cuts, number of events, and significance of signal with 3000/fb of data at the 13 TeV LHC. We have defined $\mathcal{S}\equiv S\,/\,\sqrt{B+\sigma_{bg}^2 B^2}$\,. See text for details.}
\label{tab:resultsssdl}
\end{center}
\end{table}

\vskip0.3cm
\noindent\textit{Multileptons search}
\vskip0.3cm

Next, we mirror the CMS search for a signal in multileptons (Signal Region (SR) 14, ``off-$Z$" analysis as defined in \cite{CMS-PAS-SUS-16-003}) by imposing the following set of requirements on the generated event sample (henceforth referred to as ``$\geq3l$offZ"): 
\begin{itemize}
\item Require three or more electrons or muons with $p_T \,\geq \, 20,15,10$ GeV. 
\item Require two or more jets, $N_{jets}\geq 2$.
\item Impose a $Z$ veto: reject events with opposite sign, same-flavor dilepton pairs with an invariant mass between 76 and 106 GeV.
\item Constrain the missing transverse energy to $50\,\leq\,\missET$\,$\leq\,300$ GeV.
\item Require $H_T\,\geq\,600$ GeV.
\end{itemize}

These cuts are not optimized for the signal, but they allow a robust determination of the background as determined by the experiment (which includes a not insignificant contribution from $t \bar{t}$ + fake leptons, which is difficult to estimate via naive simulation).  The expected number of background events is again taken from the CMS analysis \cite{CMS-PAS-SUS-16-003} and scaled up to 3000 fb$^{-1}$ of data; the paper also lists a detailed breakdown of individual background contributions (see SR14 ``off-$Z$" entries in Table 3 in that paper).

We improve on this CMS search strategy by further imposing the following additional requirements:
\begin{itemize}
\item Reject events with no $b$-jets. This is particularly effective in suppressing  the significant $WZ$ background (see SR14 entry, Table 3 in \cite{CMS-PAS-SUS-16-003}).
\item Require $140\,\leq$ $\missET$\,$\leq\,300$ GeV instead of $50\,\leq$ $\missET$\,$\leq\,300$ GeV.
\end{itemize}

\begin{table}[t]
\begin{center}
\begin{tabular}{|l||c|c|c|c|c|c|c|} \hline
~~Search ~ & ~$\sigma_{\text{prod}}$/fb~ & efficiency($\epsilon$)& no. of signal & background & ~~$\mathcal{S}$~~&~~$\mathcal{S}$~~&~~$\mathcal{S}$~~\\
~ & ~ & ~($\times 10^{-4}$ ) & ~events (S)~ & ~events (B)~ & $\sigma_{bg}=0$ & $\sigma_{bg}=0.1$ & $\sigma_{bg}=0.3$ \\
\hline\hline
~$\geq 3l$offZ~ & - & - & 102 & 297 & 5.9 & 3.0 & 1.1\\
\hline
~$\sbot$ contribution & 14 & 19 & 81 & - & - & - & -\\
~$\sth$ contribution & 5 & 14 & 21 & - & -& - & -\\
\hline
\end{tabular}
\caption{Multileptons search: Efficiency of cuts, number of events, and significance of signal with 3000/fb of data at the 13 TeV LHC. We have defined $\mathcal{S}\equiv S\,/\,\sqrt{B+\sigma_{bg}^2 B^2}$\,. See text for details.}
\label{tab:results1}
\end{center}
\end{table}

To estimate the background suppression from the $b$-tag requirement, we look at how individual background contributions drop when this requirement is imposed in the $50\,\leq$ $\missET$\,$\leq\,300$ GeV, $60\,\leq H_T \leq\,600$ GeV region -- this information is readily available in the CMS analysis \cite{CMS-PAS-SUS-16-003} (see Table 3, SR 1-12; see Table 1 for their definitions). To estimate the effect of the stronger $\missET$ cut, we simulate the SM $t\bar{t}Z/h$ background (one of the major backgrounds for this signal), observe how it falls for increasing $\missET$, and make the simplifying assumption that all  SM backgrounds scale in the same manner (as noted above, other large backgrounds include $t \bar{t}$ + jets with a fake lepton, which are also expected to fall at large $\missET$ \cite{CMS-PAS-SUS-16-003}.)

The resulting efficiencies, number of events, and signal significance are presented in Table\,\ref{tab:results1}. A $3\sigma$ significance appears possible with improvements in systematic uncertainty in background to $\sigma_{bg}=0.1$, and further reducing it could even enable a $5\sigma$ discovery. This search strategy is therefore slightly more efficient than the SSDL analysis in extracting the signal, although a larger fraction of the signal now comes from the heavier stop.  

\vskip0.3cm
\noindent\textit{Heavier stop search}
\vskip0.3cm

\begin{figure}[t]
\centering
\includegraphics[width=0.5\linewidth]{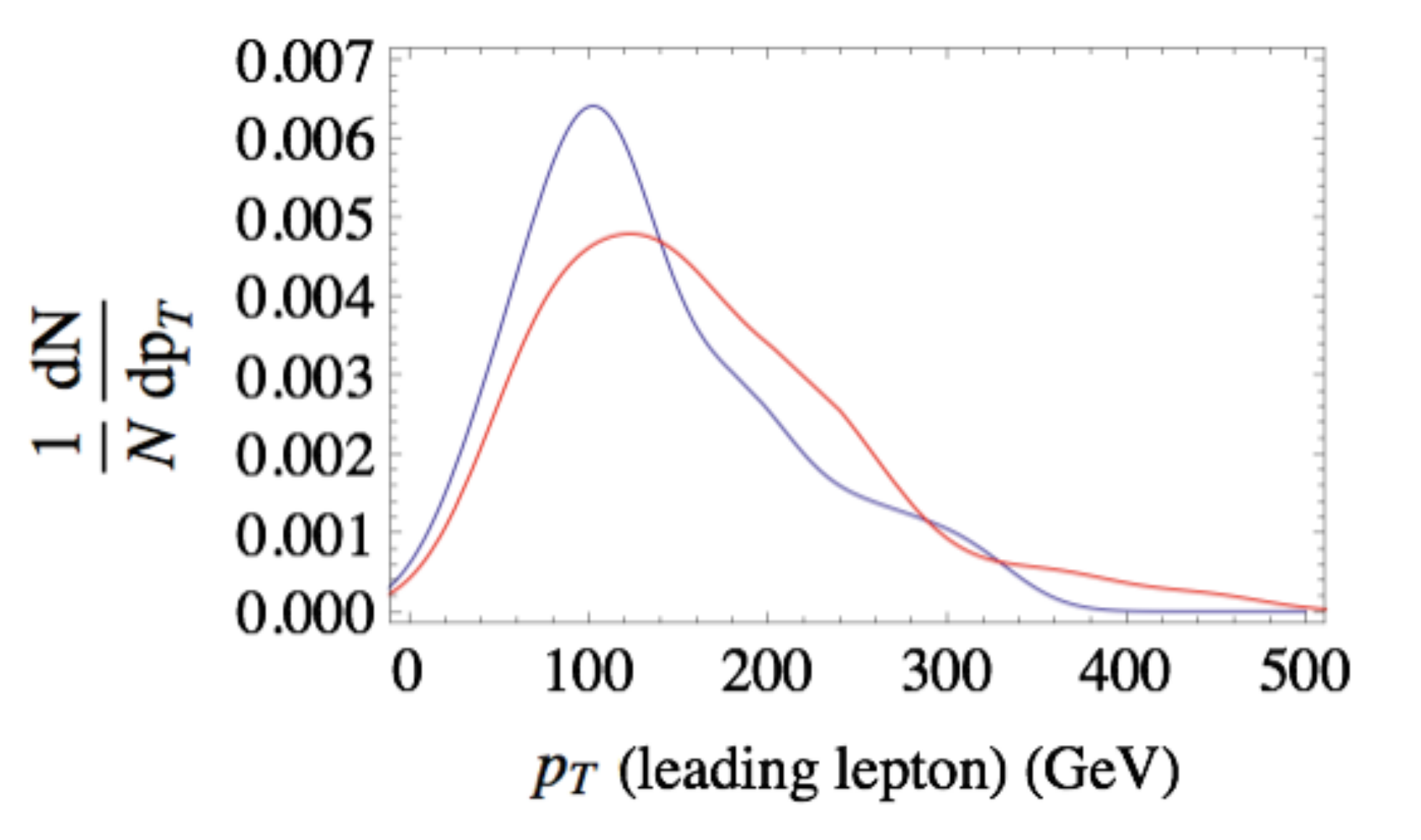}
\caption{Comparison of hardest lepton $p_T$ spectra (normalized) for two different mass spectra with mass splittings $\msbot-\mstl\,\approx\,240$ GeV (red curve) and $\msbot-\mstl\,\approx\,170$ GeV (blue curve).}
\label{fig:pTcomp}
\end{figure}

As discussed earlier, a light sbottom discovery with a large $\Delta m_{\sbot\stl}$ in the MSSM allows us to predict a TeV scale $\sth$. The large mass splitting can be established, for instance, by looking at the $p_T$ spectrum of the hardest lepton. To motivate that it is possible to draw such conclusions, in Fig.\,\ref{fig:pTcomp} we plot, in red, the (normalized) $p_T$ spectrum of the hardest lepton in the signal events in the $\geq 3l$offZ(II) analysis above, which corresponds to $\Delta m_{\sbot\stl}\,
\approx\,240$ GeV for our benchmark point. For comparison, we also plot, in blue, the corresponding spectrum for $\Delta m_{\sbot\stl}\,\approx\,170$ GeV. The red spectrum is broader and has a stronger tail (above $\sim 330$ GeV). By making use of such features, it is plausible that the $\Delta m_{\sbot\stl}$ splitting can be determined to within $100$ GeV. For our benchmark scenario, this would enable us to predict $\msth\,\lsim\,1.2$ TeV. Moreover, this also enables us to deduce that there is significant mixing between the stops (see Fig.\,\ref{fig:mixingsplitting}), and thus Br($\sth\rightarrow\stl Z$) should be significant. The next step, therefore, is to devise a search strategy for such a $\sth$. 

Again, we make use of the CMS analysis as above ($\geq3l$offZ), except we now require a $Z$-reconstruction rather than a $Z$-veto in order to search for $Z$ bosons from $\sth\to\stl Z$ decays (this is defined as SR14, ``on-$Z$" analysis in the CMS paper \cite{CMS-PAS-SUS-16-003}). The background is again taken from the CMS paper \cite{CMS-PAS-SUS-16-003} (SR14, Table 4) and scaled up to 3000 fb$^{-1}$. We optimize the CMS search strategy by imposing the following additional requirements:
\begin{itemize}
\item Reject events with no $b$-tagged jets. Again, this is particularly effective in suppressing the dominant $WZ$ background (see SR14 entry, Table 4 in \cite{CMS-PAS-SUS-16-003}).
\item Require $\missET\geq\,200$ GeV instead of $50\,\leq\missET\,\leq\,300$ GeV. Since $\sth$ is significantly heavier than $\sbot$, the $\sth$ signal contains a higher $\missET$ distribution, motivating an even higher $\missET$ cut than that employed in the CMS analysis. 
\end{itemize}

We estimate the modified background contribution in the same manner as for $\geq 3l$offZ.  We use information from Table 4, SR 1-12 from \cite{CMS-PAS-SUS-16-003} to extrapolate the effects of the $b$-tag requirement on background, and simulate $t\bar{t}Z/h$  to determine the effects  of the increased $\missET$ cut, taking it to be representative of all background.\,\footnote{Note that $\missET\geq\, 300$ GeV is a part of SR15 ($\missET\geq\, 300$ GeV, $H_T\,\geq\,600$ GeV) and not SR14 in the CMS analysis \cite{CMS-PAS-SUS-16-003}.  We have appropriately scaled the background in SR15 using results of stronger $H_T$ cuts on our simulated $t\,\bar{t}\,Z/h$ sample to estimate the modified background contribution from this region.} A similar search proposal for $\sth\,\to\,\stl\,Z$ in \cite{Perelstein:2007nx} also employed a narrower cut on the $Z$-boson lepton pair invariant mass  as a strategy to suppress background, particularly the combinatoric background from $t\bar{t}$. In the CMS analysis, the $t\bar{t}$ background is claimed to be largely suppressed by the strong $\missET$ and $H_T$ cuts, hence we do not pursue this strategy in our analysis but note this could provide a further handle.

\begin{table}[t]
\begin{center}
\begin{tabular}{|l||c|c|c|c|c|c|c|} \hline
~~Search ~ & ~$\sigma_{\text{prod}}$/fb~ & efficiency($\epsilon$)& no. of signal & background & ~~$\mathcal{S}$~~&~~$\mathcal{S}$~~&~~$\mathcal{S}$~~\\
~ & ~ & ~($\times 10^{-4}$ ) & ~events (S)~ & ~events (B)~ & $\sigma_{bg}=0$ & $\sigma_{bg}=0.1$ & $\sigma_{bg}=0.3$ \\
\hline\hline
~~$\textgreater 3l$onZ~& - & -  & 152 & 432  & 7.3 & 3.2 & 1.2 \\
\hline
~~$\sth$ contribution ~& 14 & 78  & 117 & - & - & - & -\\
~~$\sbot$ contribution ~& 5 & 8 & 35 & - & 1.7 & 0.7 & 0.3\\
\hline
\end{tabular}
\caption{Number of events and significance of signal with 3000/fb of data at the 13 TeV LHC for the $\textgreater 3l$onZ search. We have defined $\mathcal{S}\equiv S\,/\,\sqrt{B+\sigma_{bg}^2 B^2}$\,.}
\label{tab:results2}
\end{center}
\end{table}

We denote the above search as ``$\geq3l$onZ", and present the resulting efficiencies, number of events, and signal significance in Table\,\ref{tab:results2}. As with the sbottom search strategies, we see that a $\sim3\sigma$ signal is possible with improvements to $\sigma_{bg}=0.1$, and a $\sim5\sigma$ discovery is possible with further improvements. For comparison, we also list the signal significance for the sbottom contribution only, which makes it clear that sbottom pollution to this signal is minimal.

%%%%%%%%%%%%%%%%%%%%%%%%%%%%%%%%
\section{Heavier Stop Multiple Decay Channels in Boosted Dibosons}
\label{sec:stop2}
%%%%%%%%%%%%%%%%%%%%%%%%%%%%%%%%

If superpartners are discovered at the LHC, the high luminosity LHC will be able to follow up with measurements in multiple channels with significant statistics. A particularly illustrative example is the decay of the heavier stop $\sth$, which can occur in multiple channels $\stl Z,\,\stl h, \,\sbot W$, and $t \chi_0$, with branching ratios determined by the stop masses and mixing angle. 

In this section, we focus on the two decays $\sth\to\stl Z$ and $\sth\to\stl h$, which give rise to boosted dibosons if the mass splitting between the two stop mass eigenstates is large. The tree-level decay widths for these two processes, in the decoupling limit in the Higgs sector are \cite{Baer:2006rs}
\beqa
\Gamma(\tilde{t}_2\rightarrow\tilde{t}_1\,Z)&=&\frac{g^2}{256\pi}\frac{m_{\tilde{t}_2}^3}{m_W^2} \sin^2{2\theta_{\tilde{t}}}\, \lambda^{3/2}(1,m_{\tilde{t}_1}^2/m_{\tilde{t}_2}^2,m_Z^2/m_{\tilde{t}_2}^2), 
\label{eq:zwidth}\\
\Gamma(\tilde{t}_2\rightarrow\tilde{t}_1\,h)&=&\frac{g^2}{256 \pi}\frac{m_{\tilde{t}_2}^3}{M_W^2}|A_h|^2\,\lambda^{3/2}(1,m_{\tilde{t}_1}^2/m_{\tilde{t}_2}^2,m_h^2/m_{\tilde{t}_2}^2),\nonumber\\
A_h&=&\frac{M_W^2}{m_{\tilde{t}_2}^2} \left(1-\frac{5}{3}\tan^2\theta_W\right) \sin{2\theta_{\tilde{t}}}+\frac{\sin{4\theta_{\tilde{t}}}}{2} \,\left(1-\frac{m_{\tilde{t}_1}^2}{m_{\tilde{t}_2}^2}\right),
\label{eq:hwidth}
\eeqa
where the phase space factor is $\lambda(a,b,c) \equiv a^2+b^2+c^2-2ab-2ac-2bc$. The ratio of the two widths is:
 \beq
R_{hZ} \equiv\frac{\Gamma(\tilde{t}_2\rightarrow\tilde{t}_1\,h)}{\Gamma(\tilde{t}_2\rightarrow\tilde{t}_1\,Z)}=\left[\left(1-\frac{m_{\tilde{t}_1}^2}{m_{\tilde{t}_2}^2}\right) \cos{2\theta_{\tilde{t}}} +\frac{m_W^2}{m_{\tilde{t}_2}^2}\left(1-\frac{5}{3}\tan^2\theta_W\right)\right]^2\approx \left(1-\frac{m_{\tilde{t}_1}^2}{m_{\tilde{t}_2}^2}\right)^2 \cos^2{2\theta_{\tilde{t}}}.
\label{eq:Rdef}
\eeq
The phase space factors effectively cancel for $\msth-\mstl\gg m_h,m_Z$. We expect many experimental uncertainties to cancel in this ratio as well. If the two stop masses are known from other measurements, this ratio offers a clean dependence on the stop mixing angle,\,\footnote{See \cite{Rolbiecki:2009hk} for similar ideas to extract the stop mixing angle from measurement of ratios of various processes when the Higgsinos are light.} enabling a check of the MSSM Higgs mass relation. It should be clarified that we are not advocating $R_{hZ}$ as the most precise measurement of the stop mixing angle, but rather as a measurement with a particularly simple dependence on an important parameter in the theory.

An important caveat is that the above expressions only hold at tree level and will be modified by loop corrections to both $\Gamma(\tilde{t}_2\rightarrow\tilde{t}_1\,h)$ and $\Gamma(\tilde{t}_2\rightarrow\tilde{t}_1\,Z)$. The loop corrections are particularly important where the tree level contributions vanish ($\theta\to 0,\pi/2,$ for both Eq.\,(\ref{eq:zwidth}),\,(\ref{eq:hwidth}); Eq.\,(\ref{eq:hwidth}) also vanishes for $\theta\to\pi/4$).  For parameters that give large $\tilde{t}_2\rightarrow\tilde{t}_1\,h$ and/or $\tilde{t}_2\rightarrow\tilde{t}_1\,Z$ branching ratios (likely necessary for measuring these signals and providing meaningful bounds on the ratio $R_{hZ}$), loop corrections are generally subdominant. Since loop corrections introduce sensitivity to other supersymmetric parameters (such as the gluino mass), for simplicity, we choose a benchmark where the loop corrections are small and everything can be treated analytically with the relations above. Using the expressions in \cite{Bartl:1998xp,Bartl:1997pb}, we estimate that loop contributions can modify $R_{hZ}$ substantially for $\theta_t\,\textless\, 0.1$ and $\theta_t\,\textgreater\, 1.5$; we therefore exclude these regions in our analysis.\footnote{Experimentally, one might be able to confirm that nature is away from these ``loop sensitive" windows, either by measurements of $r_{gg}$, see Fig.~\ref{fig:hgg}, or by the absence of $\tilde{t}_{2} \rightarrow t \chi_0$ decays, which should be present if the branching ratios to $\stl Z$ and $\stl h$ are very small.}

\subsection{Benchmark Case Study}

The masses and branching ratios for our chosen benchmark point are listed in Table\,\ref{tab:benchmarkst2}; here we computed $m_h,\,\msbot,$ and the $\sth$ branching ratios analytically using formulae listed in the previous sections. We have chosen a point with a large combined branching fraction to $Z$ and $h$ in order to maximize our signal by rendering $\sth\to\sbot W$ kinematically inaccessible. This spectrum results in a too-light Higgs in the MSSM, so a sufficiently precise measurement of such a spectrum would imply additional new physics beyond the MSSM. For this benchmark point, $R_{hZ} = 0.53$.

\begin{table}[t]
\begin{center}
\begin{tabular}{|l|} \hline
~~~$\sth:~~~\msth=994.2$ GeV,~~~ BR:~~$\stl Z$ 52\%,~~$\stl h$ 28\%, ~~ sin$^2\,\theta_t=0.988$~~~~~\\
~~~$\sbot:~~~\msbot=977.5$ GeV,~~~ decays dominantly  to $\stl W$ \\
~~~$\stl:~~~\mstl=486.0$ GeV,~~~ decays dominantly  to $c \chi_0$ \\
~~~$\chi_0:~~~m_\chi=406.0$ GeV,~~~LSP\\
~~~$h:~~~m_h=109.2$ GeV\\
\hline
\end{tabular}
\caption{Mass spectrum and branching ratios for MSSM benchmark point. Note that while the stop and sbottom mass parameters are consistent with all current experimental constraints, the correct Higgs mass is not produced in this MSSM scenario, signaling the need for additional physics beyond the MSSM.}
\label{tab:benchmarkst2}
\end{center}
\end{table}

Measuring the ratio $R_{hZ}$ with reasonable precision requires high statistics, motivating searches for the boosted $Z$ and $h$ bosons in their dominant (hadronic) decay channels rather than the cleaner decays into leptons or photons. The prospect of probing such signals by reconstructing the boosted dibosons via fat jets was studied in Ref.\,\cite{Ghosh:2013qga}, which found that a $\sim 4 - 5\sigma$ discovery of a TeV scale $\msth$ was possible with 100 fb$^{-1}$ of data at the 14 TeV LHC with a combined diboson signal from $\tilde{t}_2 \tilde{t}_2 \rightarrow  \tilde{t}_1 \tilde{t}_1$  + ($hZ$, $ZZ$, and $hh$). For these channels, Ref.\,\cite{Ghosh:2013qga} estimates a total background cross section (after cuts) of 0.16 fb, dominated by events with two W bosons.  With a relatively narrow invariant jet mass window for Higgs boson candidates as in \cite{Ghosh:2013qga}, which can further be augmented by jet charge, we estimate that the probability of a ``$W$-jet" faking a ``Higgs jet" is very small, likely $<1\%$ (see Fig.\,2(c) in Ref.\,\cite{Aad:2015eax}). Scaling the backgrounds from \cite{Ghosh:2013qga} by this ``mistag" probability, we expect the SM backgrounds to be negligible for the $hZ$ and $hh$ channels.  Incidentally, we also expect the probability for a $Z$ jet to fake a Higgs jet to be small. $R_{hZ}$ can thus be determined in an essentially background-free environment by considering events with at least one Higgs jet:\footnote{This strategy also avoids the possibility of contamination from $\sbot$ decays with $\msbot\sim\msth$, where the fat jets from $W$ bosons from $\sbot\to\stl W$ can be misinterpreted as $Z$ bosons from $\sth\to\stl Z$. The $\sbot$ contribution was not considered in \cite{Ghosh:2013qga}.}
\bea
R_{hZ} \sim  \,\frac{2\,n_{hh}}{n_{Zh}}\,,
\eea
where $n_{ab}$ denotes the number of signal events where the two boosted dibosons are 
tagged as $a$ and $b$. The error in the calculated ratio $R^{i}_{j} =n_i/n_j$ is
\beq
\Delta R^i_j=\sqrt{\left(\frac{\Delta n_i}{n_j}\right)^2+\left(\frac{n_i \Delta n_j}{n_j^2}\right)^2}.
\eeq
Since the associated backgrounds are negligible, we estimate $\Delta n_{i}  = \sqrt{n_{i}}$.

Our benchmark point is similar those in Ref.\,\cite{Ghosh:2013qga} in terms of mass spectra and branching ratios into various decay channels. This allows for a straightforward extrapolation of the results of this study. We extract the signal efficiency from this paper and apply it to our benchmark point, applying a modest correction for the branching ratios. With the simplifying assumption that this analysis is equally efficient in extracting $Z$ and $h$ events (likely approximately true  given the nearly identical branching ratios to fully hadronic final states), we estimate an overall (efficiency $\times$ BR) of $\epsilon_{hZ} \approx 5.9 \times10^{-3}$ and $\epsilon_{hh} \approx 1.6 \times 10^{-3}$. The interested reader is referred to Ref.\,\cite{Ghosh:2013qga} for details of the analysis.
 
The resulting number of events and the corresponding uncertainty on $\Delta R_{hZ}$ for 3000 fb$^{-1}$ of data at the 14 TeV LHC are:
\beq
~~n_{hh}=47,~~~~n_{Zh}=176,~~~~\Delta R_{hZ}=.07
\eeq
For our benchmark point, $R_{hZ}$ can thus be measured to within $\sim12\%$. Whether the fat jet analyses of the type employed here can remain effective in the high luminosity environment is a question for further study.

\begin{figure}[t]
\centering
\includegraphics[width=0.5\linewidth]{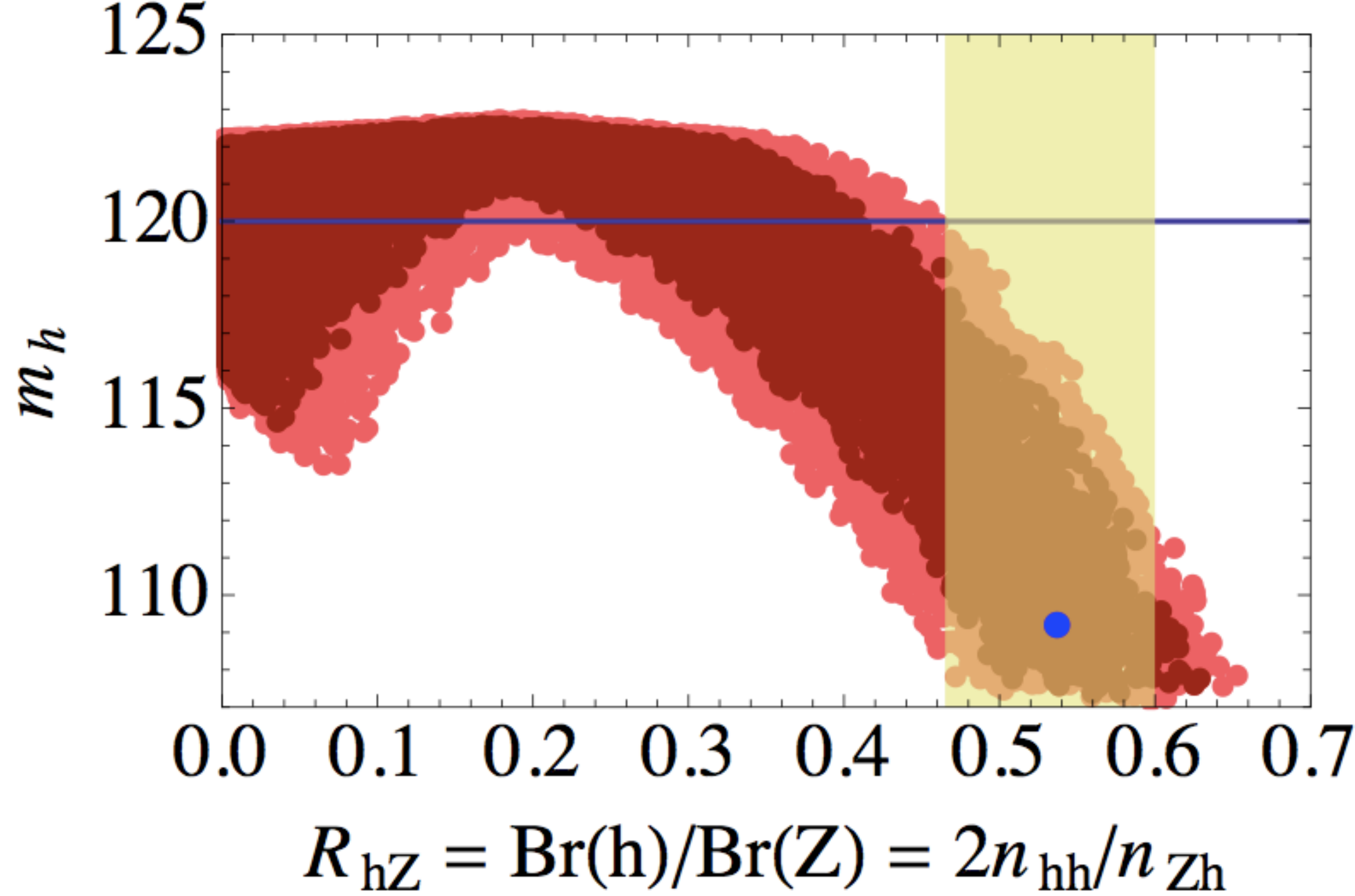}
\caption{MSSM Higgs mass as a function of $R_{hZ}$. The horizontal blue line denotes $m_h=120$ GeV, the cutoff below which the Higgs mass is taken to be inconsistent with the MSSM. Light (dark) red bands correspond to $440\leq\mstl\leq520$\,GeV and $930\leq\msth\leq1030$\,GeV ($450\leq\mstl\leq510$\,GeV and $945\leq\msth\leq1015$\,GeV). The blue dot denotes the benchmark point in our analysis. The golden band shows uncertainties in the calculated value of $R_{hZ}$ with 3000 fb$^{-1}$ of data.}
\label{fig:hZratio}
\end{figure}

Next, we discuss how this measurement can shed light on the Higgs boson mass relation and the validity of the MSSM. This requires some knowledge of the two stop masses, hence we assume that $\mstl$ has been measured to lie in the range $486\pm 40$ GeV from monojet or charm-tagged events, while $\msth$ is known to fall in the $994.2\pm 50$ GeV range from various measurements (such as by combining the knowledge of $\mstl$ with information on $p_T(Z)$  in $\sth\to\stl Z$ events). The MSSM Higgs mass can then be calculated as a function of $\theta_t$ using Eq.\,(\ref{eq:higgsmass}), which can be converted to a function of the ratio $R_{hZ}$ using Eq.\,(\ref{eq:Rdef}). We plot this dependence in Fig.\,\ref{fig:hZratio} in the broad red band for the above stop mass windows. For comparison, the narrower, darker red band corresponds to $486\pm 30$ GeV and $994.2\pm 35$ GeV for the lighter and heavier stop masses respectively, and illustrates how the uncertainty in the Higgs mass decreases with better knowledge of the stop masses.

Recall that the Higgs mass is small for vanishing stop mixing $\theta_t\to 0,\,\pi/2$, which corresponds to $R_{hZ}\sim \cos^2{2\theta_t}$ approaching 1. On the other hand, achieving the correct Higgs mass with sub-TeV stops requires large stop mixing, which correlates with a smaller value of $R_{hZ}$. The trend in Fig.\,\ref{fig:hZratio} is consistent with these observations. Thus, an inferred value of $R_{hZ}$ above some cutoff value $R_0$ ($\approx 0.45$ in this case) is incompatible with the MSSM Higgs mass relation. Such an observation would rule out the MSSM, pointing to the need for additional contributions to the Higgs mass. In Fig.\,\ref{fig:hZratio}, in the golden band we show the uncertainty in the calculated value of $R_{hZ}$ for our benchmark point. Under our assumptions, exclusion of the MSSM region is borderline; however, the MSSM can be clearly excluded with either better measurements of the stop masses (darker red band) or with an improved analysis with better signal efficiency (recall that here we simply used the efficiency from the analysis in Ref.\,\cite{Ghosh:2013qga}). This benchmark study serves as a proof of concept that measurements of the two decay channels $\sth\to\stl Z$ and $\sth\to\stl h$  can be used as a consistency check of the Higgs mass and possibly rule out the MSSM. 

We conclude this section with a few miscellaneous comments. With approximate knowledge of the two stop masses, requiring that $r_{gg}$ remain consistent with observations also constrains the stop mixing angle. For our benchmark point (with the narrower stop mass windows discussed above), we find that $r_{gg}$ measurements at the 3000 fb$^{-1}$ LHC can constrain $R_{hZ}$ to $0.2\,\textless\,R_{hZ}\,\textless\,0.6$, therefore providing complementary handles on the Higgs mass relation. Likewise, if the sbottom has not already been discovered, the above measurements can also be used to predict the mass and decay channels of the sbottom, aiding in its discovery. For our benchmark scenario, the sbottom is degenerate with $\sth$ and decays almost exclusively to $\sbot\to\stl W$, with $\mstl$ decaying as $\stl\to c\chi_0$. Dedicated searches optimized towards accepting $W$ jets instead of $Z$ jets might prove fruitful in discovering such a sbottom.

%%%%%%%%%%%%%%%%%%%%%%%%%%%%%%%%
\section{Summary}
\label{sec:summary}

In this paper, we investigated the implications of a third generation squark signal discovery at the LHC. It is possible to make use of the relation between the stop sector and the Higgs boson mass in the MSSM in a wide variety of scenarios to test the consistency of the MSSM and predict the masses and decay channels of other superpartners, therefore offering clear subsequent targets for the LHC. We elaborated these ideas with studies in three distinct scenarios:

\begin{itemize}
\item For a light ($450\leq \mstl\leq 600$ GeV) stop, constraints on the Higgs-gluon-gluon coupling strongly limit the stop sector parameters, which can be translated into bounds on the $\sbot$ and $\sth$ masses, leading to interesting patterns in the MSSM. For instance, the MSSM Higgs mass relation forces $\Delta m_{\sbot\stl}\,\textless\, m_W$ for $\mstl\lsim500$\,GeV if $\stl$ is left-handed, which is incompatible with current LHC constraints. Likewise, scenarios involving $\sbot\rightarrow\stl W$ or $\sth\rightarrow\sbot W$ require significant stop mixing, and consistency with the Higgs mass in the MSSM leads to the prediction $\msth\,\lsim1.2$ TeV. 

\item In the event of a sbottom signal in same-sign dileptons or multileptons+jets+$\missET$ from $\sbot\rightarrow\stl W$, the correlation between $\Delta m_{\sbot\stl}$ and $\msth$ in the MSSM can be used to predict the $\sth$ mass and decay channels, thereby aiding $\sth$ searches at the LHC. 

\item For fixed stop masses, the ratio of $\sth\rightarrow \stl Z$ and $\sth\rightarrow \stl h$ decay widths is determined by the stop mixing angle, and measuring this ratio with sufficient precision can test the MSSM Higgs mass relation and therefore check the validity of the MSSM. We illustrated the plausibility of this scenario with a case study of the reconstruction of the boosted dibosons from fat jets at the high luminosity LHC with  3/ab of data. 
\end{itemize} 

These examples do not cover the full range of signals or spectra that are possible in the MSSM, and it might also be interesting to perform similar studies focusing on signals involving, $\textit{e.g.}$, light Higgsinos or the gluino. Likewise, we employed several approximations in our studies.  Higher precision calculations and more careful simulations would be warranted should a relevant signal actually be discovered at the LHC. Nevertheless, the scenarios we studied here illustrate the power and applicability of the Higgs mass relation in unravelling the supersymmetric sector. 

\acknowledgments
We thank Bob Zheng for valuable discussions and collaboration in the early stages of this project. We also acknowledge  helpful discussions with Haipeng An, Diptimoy Ghosh, Nausheen Shah, and Carlos Wagner. This work is supported by the U.S. Department of Energy, Office of Science, under grant DE-SC0007859. BS also acknowledges support from the University of Cincinnati. This work was performed in part at the Aspen Center for Physics, which is supported by National Science Foundation grant PHY-1066293.

%%%%%%%%%%%%%%%%%%%%%%%%%%%%%%%%
\bibliography{stopbib}

\end{document}